\documentclass[journal]{IEEEtran}

\usepackage{graphicx}
\usepackage{caption}
\usepackage{subcaption}  
\usepackage{amsmath}
\usepackage{color}
\usepackage{comment}
\usepackage{url}
\usepackage{tabularx} 
\usepackage{array}
\usepackage{booktabs}
\usepackage{hyperref}
\usepackage{enumitem}
\usepackage{chemmacros}
\usepackage{balance}

\title{Artificial Intelligence for Atmospheric Sciences: \\ A Research Roadmap}

\author{Martha Arbayani Zaidan, \IEEEmembership{Senior Member, IEEE}, Naser Hossein Motlagh, \IEEEmembership{Senior Member, IEEE}, \\ Petteri Nurmi, Tareq Hussein, Markku Kulmala, Tuukka Pet\"aj\"a, Sasu Tarkoma, \IEEEmembership{Senior Member, IEEE}  
\thanks{
This project is supported by the Research Council of Finland through the University Profiling funding initiative InterEarth (Grant No. 353218) and an Academy Research Fellowship (Grant No. 355330). In addition, financial supports through the Research Council of Finland via the Atmosphere, Climate Competence Center (project numbers 337549, 357902, 359340) and via Research Infrastructure of Institute for Atmospheric \& Earth System Research  - INAR RI (367739) are gratefully acknowledged.}
\thanks{M.A. Zaidan is with Department of Computer Science and Institute for Atmospheric Science and Earth System Research (INAR), University of Helsinki, Finland (e-mail: martha.zaidan@helsinki.fi).} 
\thanks{N. H. Motlagh is with Department of Computer Science, University of Helsinki, Finland (e-mail: naser.motlagh@helsinki.fi).} 
\thanks{P. Nurmi is with Department of Computer Science and Helsinki Institute of Sustainability Science (HELSUS), University of Helsinki, Finland (e-mail: petteri.nurmi@helsinki.fi)} 
\thanks{T. Hussein is with Institute for Atmospheric Science and Earth System Research (INAR), University of Helsinki, Finland (e-mail: tareq.hussein@helsinki.fi).}
\thanks{M. Kulmala is with Institute for Atmospheric Science and Earth System Research (INAR), University of Helsinki, Finland (e-mail: markku.kulmala@helsinki.fi).} 
\thanks{T. Petäjä is with Institute for Atmospheric Science and Earth System Research (INAR), University of Helsinki, Finland (e-mail: tuukka.petaja@helsinki.fi).} 
\thanks{S. Takoma is with Department of Computer Science, University of Helsinki, Finland (e-mail: sasu.tarkoma@helsinki.fi).}  
\thanks{(Corresponding authors: M.A. Zaidan.)}
}

\begin{document}


\maketitle


\begin{abstract}
Atmospheric sciences are crucial for understanding environmental phenomena ranging from air quality to extreme weather events, and climate change. Recent breakthroughs in sensing, communication, computing, and Artificial Intelligence (AI) have significantly advanced atmospheric sciences, enabling the generation of vast amounts of data through long-term Earth observations and providing powerful tools for analyzing atmospheric phenomena and predicting natural disasters.  This paper contributes a critical interdisciplinary overview that bridges the fields of atmospheric science and computer science, highlighting the transformative potential of AI in atmospheric research. We identify key challenges associated with integrating AI into atmospheric research, including issues related to big data and infrastructure, and provide a detailed research roadmap that addresses both current and emerging challenges. 
\end{abstract}

\section{Introduction} \label{sec:introduction}

\IEEEPARstart{U}{n}derstanding the atmosphere and environment is crucial for sustaining life on Earth. Atmospheric and environmental sciences play a key role in mitigating climate change, predict natural disasters, and ensuring the health of humans and ecosystems~\cite{fu2021research}. This importance is highlighted by the increasing frequency of severe weather events and environmental degradation, which together pose a looming threat to global stability~\cite{abbass2022review}. For instance, November 2024 was recorded as the second-warmest November globally, with temperatures averaging 1.62°C above pre-industrial levels. Similarly, devastating floods in Spain in 2024 were intensified by excessive heat in the Mediterranean and Atlantic Oceans, exemplifying the interconnectedness of rising sea temperatures and extreme weather~\cite{wmo2024spainflood}. These and numerous other related events highlight the accelerating effects of climate change and reinforce the need for immediate action~\cite{copernicus2024warmest}. Accurate monitoring of atmospheric conditions, from local weather patterns to global climate models, is vital for addressing these challenges as it enables policymakers, scientists, and industry leaders to make informed decisions that impact both the environment and society~\cite{Regan2023towards}.

Recent breakthroughs in sensing, communication, computing, and Artificial Intelligence (AI) have significantly advanced atmospheric sciences by facilitating the generation of extensive datasets through long-term Earth observations and providing powerful tools to better understand atmospheric phenomena and predict events such as natural disasters. However, these advancements also present new challenges. Firstly, the rapid growth of big data from ground measurements, satellite observations, laboratory experiments, and Internet of Things (IoT) sensor networks is overwhelming conventional analysis tools~\cite{vaduva2020scientific,vance2024big}. Satellite remote sensing alone produces vast amounts of high-dimensional data, while numerical simulations of weather patterns and environmental changes introduce computationally intensive datasets~\cite{zhang2022deep}. Similarly, the proliferation of low-cost IoT sensor networks has dramatically increased the availability of real-time atmospheric data, complicating data management and analysis~\cite{talebkhah2021iot,hajjaji2021big,salam2024internet}.  Secondly,  maintaining the extensive infrastructure required for various sensing, communication, and computing platforms presents its own challenges. Ensuring the accuracy, reliability, and availability of these systems requires continuous calibration, data synchronization, and fault management~\cite{poupry2023data}, particularly in remote or extreme environments, where wear and calibration drift can degrade data quality~\cite{sokhi2021advances}.


Addressing these challenges requires advanced computing resources and innovative Artificial Intelligence (AI) methodologies. High Performance Computing (HPC) is essential for processing, storing, and analyzing vast datasets in real-time, whether from high-dimensional satellite imagery, complex simulations, or large-scale sensor networks~\cite{zhang2022state}. AI-driven approaches are revolutionizing atmospheric infrastructure management by automating critical tasks such as sensor calibration, anomaly detection, and predictive maintenance~\cite{tuia2021toward}. These innovations enhance the efficiency and reliability of distributed monitoring systems while reducing operational costs and human intervention~\cite{zaidan2022dense}. In parallel, AI, defined broadly as the simulation of human intelligence processes by machines, is revolutionizing many scientific fields \cite{ahmed2022artificial}, including atmospheric sciences. AI encompasses a range of techniques, from machine learning algorithms to deep learning methods capable of identifying complex relationships in large datasets, making these AI-driven methods particularly well-suited for atmospheric research. These methods are particularly well-suited for processing the highly varied and voluminous data that characterize atmospheric and environmental research. For example, by applying AI, researchers can optimize weather prediction models, improve climate simulations, and derive actionable insights from satellite imagery or sensor data~\cite{dueben2022challenges}.

This paper contributes a critical interdisciplinary overview that bridges the fields of atmospheric science and computer science, highlighting the transformative potential of AI in atmospheric research. We identify and analyze key challenges posed by big data and extensive infrastructure, providing a comprehensive assessment of current AI-driven methodologies and solutions that enhance data analysis and predictive modeling. In addition, we outline a research roadmap that emphasizes emerging trends such as edge computing, AI-accelerated modeling, and interdisciplinary collaboration, fostering synergies between atmospheric science and computational technologies. By clarifying these intersections and offering practical pathways forward, this work aims to catalyze further advancements in environmental monitoring, disaster prediction, and climate change mitigation efforts, contributing to more sustainable interactions with our planet's atmosphere.

The paper is organized as follows: in Section \ref{sec:infrastructure}, we review the state-of-the-art atmospheric research infrastructure, identify existing challenges, and explore AI-driven solutions that enhance data analysis and predictive modeling. Section \ref{sec:AI} explores various branches of atmospheric sciences, analyzing current practices, challenges, and potential advancements in AI for areas such as air quality monitoring, operational meteorology, satellite remote sensing, and Earth system modeling. This section highlights key applications of AI while providing context on related research fields in environmental monitoring. In Section \ref{sec:roadmap}, we present a research roadmap that outlines future directions for AI in atmospheric sciences, emphasizing emerging trends and opportunities for interdisciplinary collaboration. Finally, Section \ref{sec:conclusion} summarizes the key insights from the paper, reinforcing the transformative role of AI in advancing atmospheric sciences.

\section{Infrastructure for Atmospheric Monitoring and Analytics} 
\label{sec:infrastructure}

\begin{figure}[h]
    \centering    \includegraphics[width=0.9\linewidth]{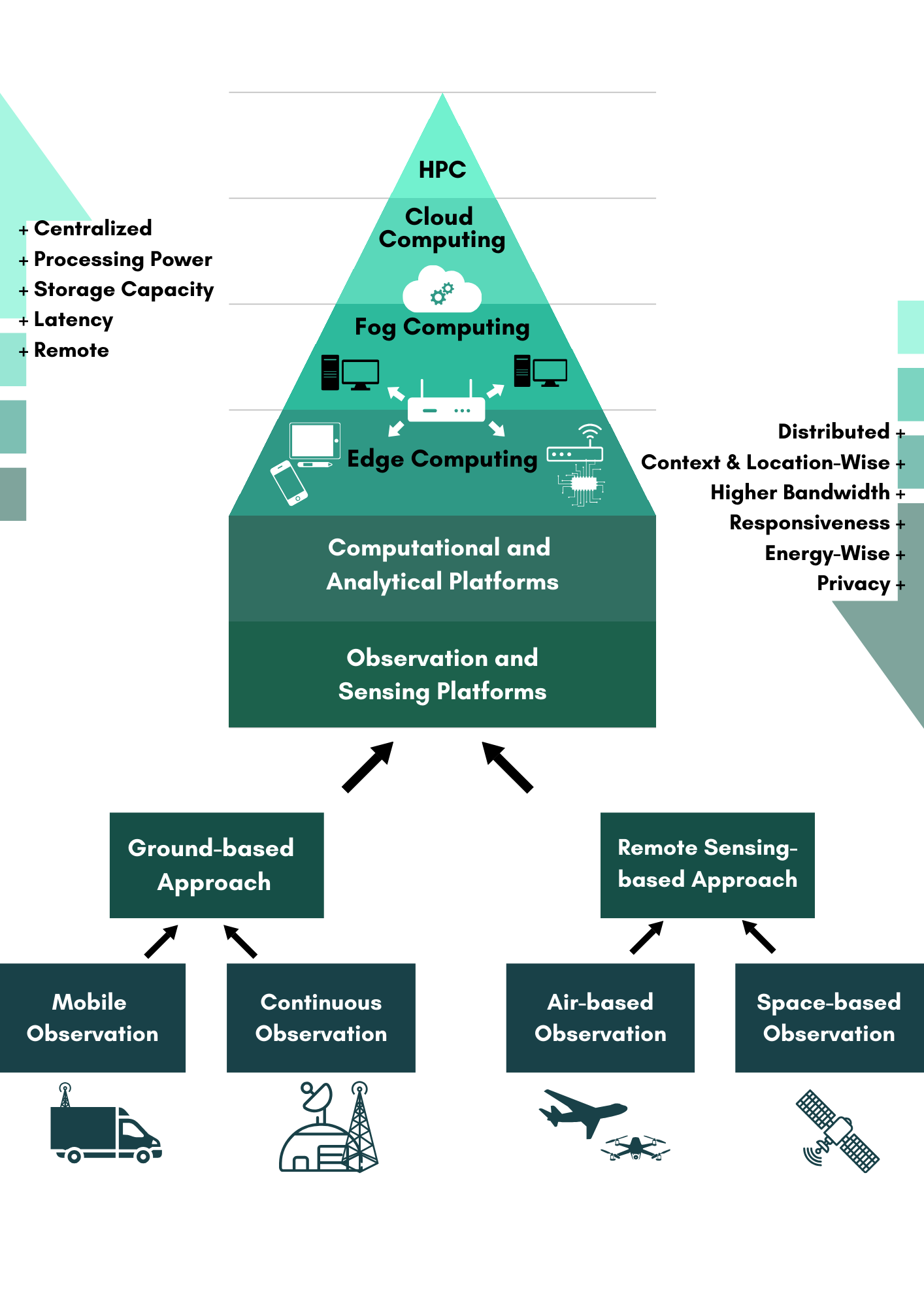}
    \caption{Atmospheric science infrastructure consists of (A) Observation and Sensing Systems and (B) Computational and Analytical Platforms.}
    \label{fig:hpc-cloud_edge_fog}
\end{figure}

Reliable infrastructure is crucial for the advancement of atmospheric sciences, as it facilitates the seamless acquisition, analysis, and dissemination of essential environmental data~\cite{petzold2024opinion}. As illustrated in Figure~\ref{fig:hpc-cloud_edge_fog}, this infrastructure can be broadly categorized into two key components:\begin{enumerate}[label=(\Alph*)]
    \item \textbf{Observation and Sensing Systems}, which include ground-based research stations, in-situ monitoring instruments, and satellite networks \cite{kulmala2018build}. These systems are designed to collect high-quality, continuous, and spatially diverse data essential for atmospheric studies and to provide the inputs needed for AI techniques.
    \item \textbf{Computational and Analytical Platforms}, such as high-performance computing (HPC) systems, cloud-based platforms, and advanced data analytics tools~\cite{li2016a2ci}. These platforms enable the processing of large-scale datasets, complex simulations, and the integration of AI-driven methodologies into atmoshperic sciences.
\end{enumerate}

Access to sufficiently sophisticated sensing and computing infrastructures is indispensable for advancing of atmospheric research, particularly in the AI era, where studies require substantial computational power and extensive datasets~\cite{bluestein2022atmospheric}. These infrastructures provide decentralized resources for both data collection and computation. The widespread availability of sensing systems enables fine-grained, high-resolution measurements, ensuring a comprehensive coverage of atmospheric phenomena. Meanwhile, computational resources support low-latency analysis of large atmospheric datasets. We next discuss these two categories of infrastructure and their transformative potential for atmospheric sciences.


\subsection{Observation and Sensing Systems}

As illustrated in Figure \ref{fig:hpc-cloud_edge_fog} (the lower part), observation and sensing systems can be categorized into ground-based and remote sensing approaches. 

\textbf{Ground-based observations} are primarily located near the Earth's surface and include both mobile and continuous observation platforms. Mobile observation platforms include vehicles, pedestrians, or other mobile units that transport data collection units, as well as fixed sensors that may be relocated periodically. These extend measurements from a single point to a trajectory with some degree of spatial resolution. However, they are significantly influenced by interference factors. For example, vehicular trajectories are affected by the driving state and road conditions. In contrast, continuous monitoring platforms typically offer lower spatial resolution, but provide stable and long-term monitoring capabilities.

Ground-based research stations, as examples of continuous observation platforms, are a vital component of atmospheric research infrastructure. For instance, the SMEAR (Stations for Measuring Ecosystem–Atmosphere Relations) stations in Finland provide continuous, long-term measurements of atmospheric, ecological and meteorological variables~\cite{kulmala2018build}. These stations generate critical data on greenhouse gas concentrations, aerosols, and other atmospheric constituents~\cite{hari2013station}, which are instrumental, e.g., for calibrating AI models for air quality sensor calibration and developing virtual sensors~\cite{zaidan2020intelligent}. 

\begin{figure}[htbp]
    \centering
    \begin{subfigure}[b]{0.45\textwidth}  
        \centering
        \includegraphics[width=\textwidth]{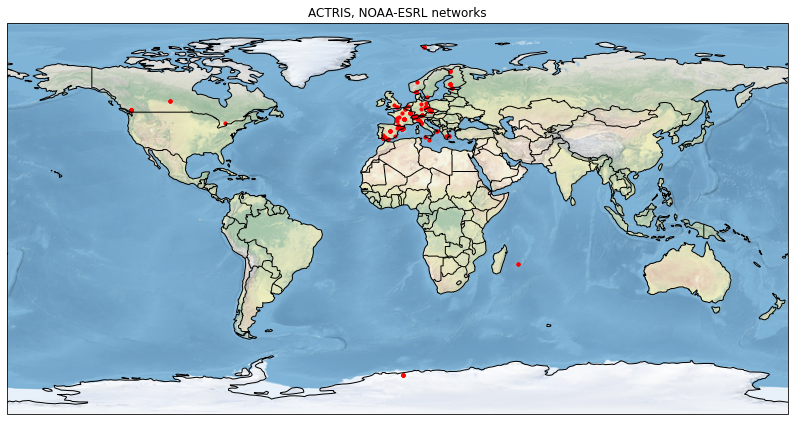}
        \caption{Aerosol,
Clouds and Trace gases Research Infrastructure Stations (ACTRIS) and NOAA Earth System Research Laboratories (ESRL) networks}  
        \label{fig:actris}
    \end{subfigure}
    \vspace{0.5cm} 

    \begin{subfigure}[b]{0.45\textwidth}  
        \centering
        \includegraphics[width=\textwidth]{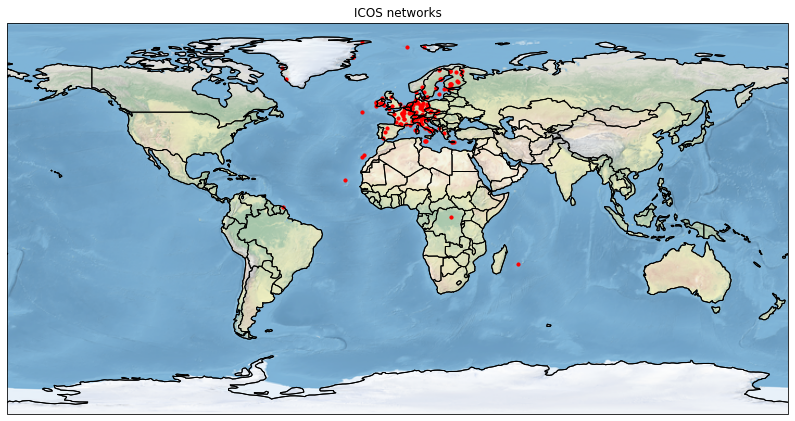}
        \caption{Integrated Carbon Observation System (ICOS) networks}  
        \label{fig:icos}
    \end{subfigure}
    \vspace{0.5cm} 

    \begin{subfigure}[b]{0.45\textwidth}  
        \centering
        \includegraphics[width=\textwidth]{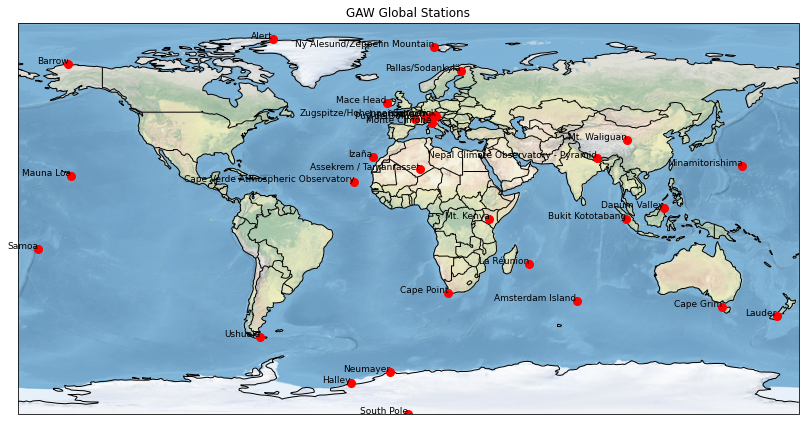}
        \caption{Global Atmosphere Watch (GAW) Global stations}  
        \label{fig:gaw}
    \end{subfigure}
    \caption{Examples of the locations of atmospheric observation infrastructure, with station data retrieved in 2024. These networks operate continuously to monitor a wide range of atmospheric variables, providing essential data for environmental research, modeling, and policy support.}
    \label{fig:three_column}
\end{figure}

Research infrastructure networks such as ACTRIS (Aerosol, Clouds and Trace Gases Research Infrastructure)~\cite{Laj2024} and ICOS (Integrated Carbon Observation System)~\cite{Heiskanen2022} provide coordinated and standardized data from a wide range of monitoring stations across Europe. Another example is the Global Atmosphere Watch (GAW) program, coordinated by the World Meteorological Organization (WMO), which focuses on building a coordinated global understanding of atmospheric composition and its changes and on helping to improve the understanding of interactions between the atmosphere, the oceans, and the biosphere. 
As shown in Figures \ref{fig:actris}, \ref{fig:icos}, and \ref{fig:gaw}, the ACTRIS, ICOS, and GAW networks span diverse geographical locations. These networks incorporate advanced and high-quality instrumentation, ensuring consistent collection of reliable and standardized data. This consistency is vital for AI-driven atmospheric research, supporting accurate and scalable atmospheric observations and simulations.

\begin{figure*}[htbp]
    \centering
    \includegraphics[width=\textwidth]{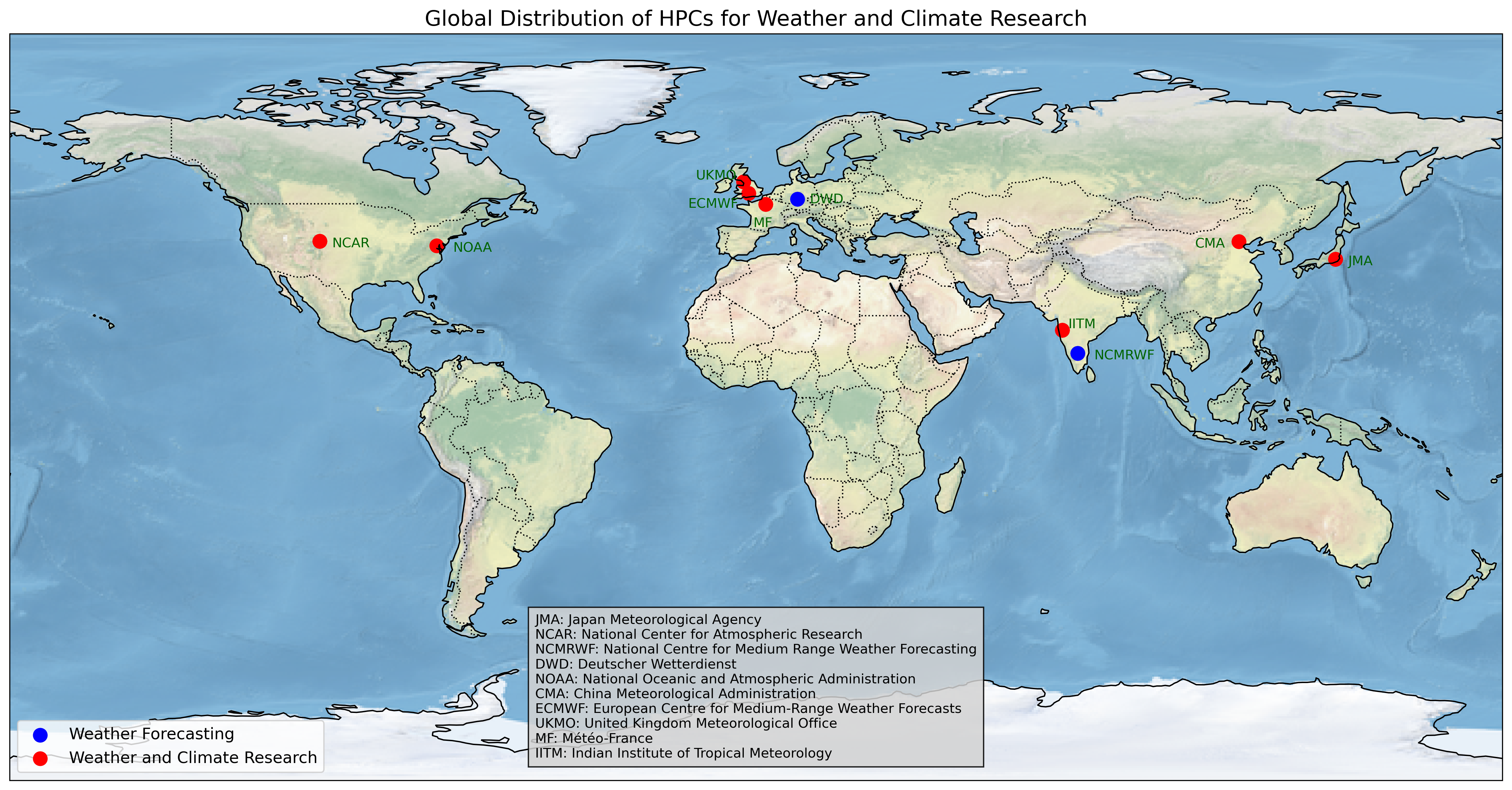}
   \caption{Locations of selected High-Performance Computing (HPC) facilities dedicated to weather forecasting (blue) and weather and climate research (red), based on data retrieved from the TOP500 list in 2024.}
    \label{fig:hpc}
\end{figure*}

\textbf{Remote sensing approaches}, in contrast, often rely on optical sensing technologies to perform atmospheric observations. These approaches can be divided into air- and space-based observation platforms.
Air-based platforms, which include aircraft, tethered balloons, and unmanned aerial vehicles (UAVs), can achieve large-scale observations by mounting instruments on these platforms. The ability to control flying altitude enables the retrieval of vertical atmospheric profiles. However, air-based platforms are also restricted by meteorological conditions.
Space-based platforms, such as satellites, have the advantage of a broad field of view, enabling the observation of large areas or even the entire globe in a short period. However, spatial resolution varies widely between satellites, some capable of sub-meter resolution and others only at the kilometer scale~\cite{zhou2023review}.
Nevertheless, satellite observations are indispensable for understanding the Earth's atmosphere~\cite{zhao2022overview}. 


\textbf{Emerging technologies} such as the Internet of Things (IoT) offer effective solutions for atmospheric observations~\cite{saeed2020around}. 
IoT-based systems integrate a network of low-cost, distributed sensors that provide real-time, high-resolution data at local scales, such as for monitoring climate~\cite{salam2024internet}, weather~\cite{zaidan2024irmaset} and air quality~\cite{hashmy2023modular}. These systems complement traditional approaches by filling observation gaps and providing data for AI-based analytics and decision-making~\cite{Motlagh2024book}. At the same time, they also face challenges related to security, privacy, and the need for robust connectivity.

One major challenge in atmospheric science is the integration of data from different sources and the need for consistent data standards across various research networks and stations~\cite{salcedo2020machine}. Ensuring data interoperability is crucial for AI models, which often require large and diverse datasets for training. Inconsistent data formats, measurement techniques, and reporting standards across different research networks can lead to challenges in data harmonization and limit the effectiveness of AI applications~\cite{yue2022towards}.

Another significant challenge pertains to the operation and maintenance of an extensive environmental monitoring infrastructure. Networks of ground-based stations, remote sensing equipment, and IoT sensors require regular calibration, data synchronization, and fault management to ensure accuracy and reliability~\cite{poupry2023data}. Maintaining this massive infrastructure not only demands considerable resources but also introduces logistical challenges, as sensors and stations deployed in remote or extreme environments often face accelerated wear and calibration drift, impacting data quality and consistency~\cite{sokhi2021advances}. 

Concurrently, modern data science approaches, particularly those based on AI and machine learning (ML), enable researchers not only to handle massive volumes of data but also to effectively manage and optimize large-scale atmospheric infrastructure~\cite{tuia2021toward}. 
Through AI-driven solutions, atmospheric scientists can automate critical tasks such as sensor calibration, anomaly detection, and system maintenance across extensive measurement networks, ensuring high data quality and reliability of distributed monitoring stations while significantly reducing operational overhead~\cite{zaidan2022dense}. By integrating advanced computing infrastructure with AI, atmospheric science can gain deeper insight into atmospheric processes and environmental changes, maintaining rigorous standards of data accuracy and enhancing infrastructure efficiency.

\subsection{Computational and Analytical Platforms}

Computational and analytical platforms are foundational components of atmospheric science infrastructure, enabling researchers to handle, process, and analyze large-scale, multidimensional datasets critical for advancing our understanding of atmospheric systems. As shown in Figure~\ref{fig:hpc-cloud_edge_fog} (the upper part), these platforms encompass high-performance computing (HPC), cloud computing, fog computing, and edge computing. 
Each serves a unique role in data processing and analytics, catering to different scales, latency requirements, and computational needs.

\textbf{High Performance Computing (HPC)} platforms provide centralized, large-scale computational resources optimized for performing highly intensive tasks~\cite{gill2024modern}, such as climate modeling~\cite{wedi2022destination}, real-time weather prediction~\cite{Michalakes2020}, and atmospheric chemistry simulations~\cite{alvanos2019accelerating}. 
HPC services are typically provided by national governments, universities, and private institutions to support scientific research and development.
These systems consist of powerful supercomputers designed to handle complex calculations and process vast datasets in parallel efficiently. 
Figure \ref{fig:hpc} shows the top 500 HPC dedicated to weather and climate research activities. 
One prominent example is the European Centre for Medium-Range Weather Forecasts (ECMWF), which leverages HPC infrastructure to perform global weather forecasts that demand high computational precision and rapid processing \cite{kukkonen2012review}. HPC platforms are particularly well-suited for tasks requiring maximum processing power and minimal latency, making them indispensable for advancing atmospheric sciences. However, their reliance on centralized infrastructure and robust connectivity poses certain limitations in accessibility and flexibility.

\textbf{Cloud Computing} offers flexible, on-demand access to computational resources and tools via remote servers, making it highly scalable and accessible~\cite{voorsluys2011introduction}.  
Unlike HPC, which is typically government-provided, cloud services are usually offered by private companies and primarily rely on large-scale data centers. However, they can also utilize distributed architectures that operate across multiple geographic regions~\cite{Monti2011}.
Cloud platforms are increasingly essential for handling large-scale computational demands of climate and environmental research, offering scalability and flexibility to meet diverse research needs. They support applications such as distributed analysis of atmospheric datasets, integration of observational and model data, and even AI-driven processes such as machine learning model training and predictive analytics~\cite{montes2020cloud}. 
For example, Google has developed a cloud computing platform, called Google Earth Engine (GEE), to effectively facilitate processing large geo data over large areas and monitoring the environment for long periods of time~\cite{amani2020google}. 
Furthermore, by leveraging adaptable on-demand resources, cloud platforms reduce the overhead of maintaining physical infrastructure, making them particularly advantageous for interdisciplinary research teams and projects requiring dynamic resource allocation~\cite{yang2017big}.


\begin{table*}[!htbp]
  \centering
    \caption{Summary of infrastructure for atmospheric monitoring and analytics}
    \label{tab:infra_challenges}
    \begin{tabularx}{\textwidth}{p{0.15\linewidth}XXX}
    \toprule
       \textbf{Infrastructure} & \textbf{State-of-the-Art} & \textbf{Challenges} & \textbf{Potential Solutions}  \\
    \midrule    
\textbf{Observation and Sensing Systems} & 
A combination of mobile, continuous, airborne, space-based, and IoT-enabled observation networks & 
Ensuring continuous operation and maintenance of extensive monitoring networks, including calibration, data synchronization, and fault management to maintain accuracy and reliability & 
AI-driven automation for sensor calibration, anomaly detection, and predictive maintenance can optimize infrastructure performance and reduce operational costs  \\
\hline
\textbf{Computational and Analytical Platforms} & 
A diverse computing ecosystem, including HPC, cloud, fog, and edge computing platforms & 
Data interoperability issues due to varying formats across research networks, along with limited HPC access in certain regions & 
International organizations like WMO and ACTRIS should standardize data formats and promote data sharing. Cloud-based platforms such as Google Earth Engine and Copernicus Climate Data Store enhance accessibility to large datasets and computational resources, enabling broader AI-driven research  \\
\bottomrule
    \end{tabularx}
\end{table*}

\textbf{Fog Computing} serves as an intermediary layer that extends cloud services closer to local edge devices, enabling faster data processing and reduced latency for time-sensitive applications. This proximity reduces latency, improves efficiency, and enhances performance for time-sensitive applications~\cite{angel2021recent}. For example, a service framework employing a distributed model averaging algorithm based on the fog computing paradigm and federated learning is proposed to facilitate accurate data fusion and real-time analysis of heterogeneous air quality data in Beijing, China. This approach successfully integrated diverse data sources to deliver high-precision insights for air quality monitoring~\cite{wang2019environmental}.

\textbf{Edge Computing} takes computation even closer to the data source than fog computing, typically directly on the sensing device or the nearby gateway. It enables ultra-low latency and minimal reliance on connectivity by processing data locally \cite{kong2022edge}. Edge platforms are best suited for applications where immediate local data processing is required to reduce latency. This also holds for many AI-driven environmental monitoring systems, such as detecting harmful gases~\cite{salameh2020end} or fire hazards~\cite{gaur2019fire}. Another example is the integration of AI algorithms, including deep learning techniques, to process hyperspectral imaging data on FPGA edge devices installed onboard unmanned aerial vehicles (UAVs). These UAVs provide a flexible and mobile solution for air quality monitoring, enhancing spatial coverage and complementing continuous monitoring stations by accessing areas that are difficult to reach with continuous monitoring stations \cite{Huang2024}.


\subsection{Challenges and Possible Solutions}

Despite the many strengths of the current research infrastructure, there are also challenges. One of the most significant challenges is the uneven accessibility of computational resources, particularly for researchers in developing countries or institutions with limited access to HPC systems. Although cloud infrastructure offers potential for overcoming accessibility barriers, disparities in internet connectivity and infrastructure still hinder these researchers, again particularly the developing countries that are the most vulnerable~\cite{arora2023building}. In addition, maintaining and upgrading existing research stations and infrastructure networks requires significant investment and coordination at the international level~\cite{pandey2022beyond}.

Advancements in research infrastructure are underway to address these challenges. Efforts to standardize data formats and improve data sharing among international networks, such as the WMO Integrated Global Observing System (WIGOS), are helping to create a more unified framework for atmospheric research \cite{pauley2022assimilation,Bowdalo2024}. Furthermore, the development of cloud-based platforms such as Google Earth Engine \cite{amani2020google} and the Copernicus Climate Data Store \cite{buontempo2022copernicus} provides easier access to vast datasets and computational resources, allowing a broader range of researchers to engage in AI-driven atmospheric studies.

To summarize, a robust and integrated research infrastructure, both for collecting observations and analyzing the measurements, is critical for advancing the state of atmospheric research. Without sufficient data, the potential benefits of AI models are hard to achieve, whereas computational infrastructure is needed to harness these models. There are many kind of systems for both purposes, ranging from  HPC systems, cloud platforms, research stations, and international monitoring networks, with all of them having their own advantages and disadvantages. To this end, it is essential to understand how these infrastructure can be best harnessed to advance the adoption of AI applications in atmospheric sciences. Continued efforts to improve data interoperability and broaden access to computational resources will further empower researchers to address complex challenges related to climate change, air quality, and atmospheric dynamics. Table~\ref{tab:infra_challenges} provides a summary of the state-of-the-art infrastructure, key challenges, and potential AI-driven solutions for both categories of infrastructure.

\section{Artificial Intelligence for Atmospheric Data Sciences} \label{sec:AI}


The infrastructure introduced in the previous section provides the foundation for leveraging AI by enabling both the collection of relevant data and the hardware on which to model and analyze it. In this section, we shift our focus to the AI models themselves, exploring various examples of how AI is being utilized across different domains of atmospheric sciences. We discuss specific applications, key challenges, and recent advancements, highlighting the transformative potential of AI in addressing contemporary environmental challenges.

\subsection{Overview of Artificial Intelligence}

This section provides an overview of Artificial Intelligence (AI) and its subsets, highlighting their tasks and methods relevant to applications in atmospheric sciences.


\subsubsection{AI methodologies}
Artificial Intelligence (AI) encompasses a broad range of technologies and methodologies that enable machines to simulate human intelligence processes, including learning, reasoning, and decision-making. In atmospheric and environmental monitoring, AI holds significant potential to revolutionize the observation, modeling, and prediction of atmospheric phenomena. As the field of atmospheric science has transitioned into the era of big data, AI has become indispensable for uncovering patterns, extracting insights, and automating decision-making processes.

\begin{center}
\begin{figure}[htbp]
    \centering  
    \includegraphics[width=0.475\textwidth]{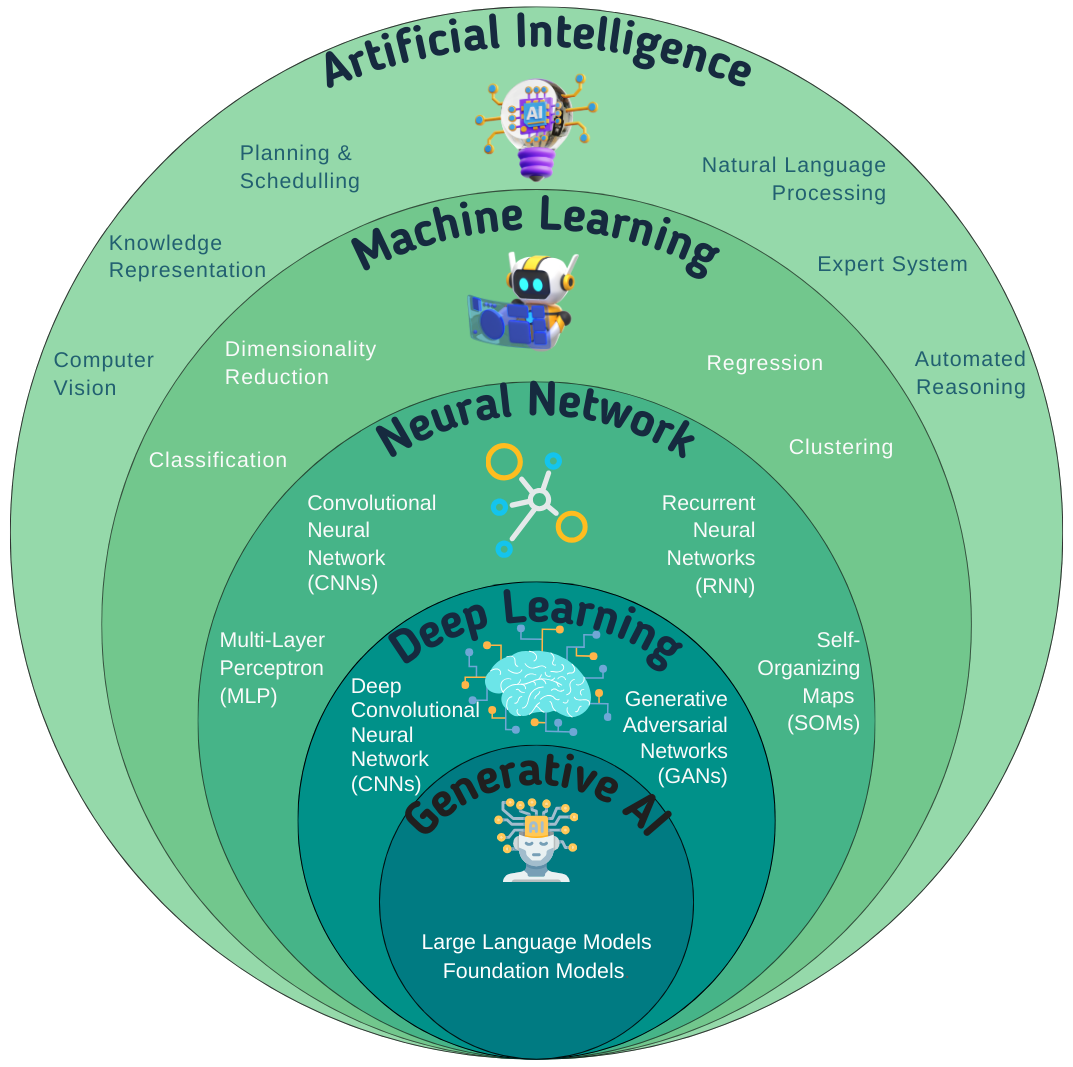}
    \caption{Artificial Intelligence with its prominent subsets}
    \label{fig:AIgroups}
\end{figure}
\end{center}

Among the various AI methodologies, Machine Learning (ML) serves as a key subset that focuses on algorithms enabling systems to learn from data and make data-driven predictions. Initially, the integration of AI in atmospheric sciences began with ML techniques, which provided data-driven alternatives to traditional numerical modeling approaches. ML algorithms are designed to learn from historical datasets and make predictions or classifications without explicitly programmed instructions. These methods have proven particularly effective for applications such as air quality forecasting, anomaly detection, and climate trend analysis~\cite{jones2017machine, wang2024howAI}.

As the volume and complexity of data grew, Deep Learning (DL), a specialized area within ML that employs multi-layered neural networks, gained prominence due to its ability to process high-dimensional, non-linear data. DL models have significantly enhanced atmospheric applications, such as improving the detection and prediction of extreme weather events~\cite{racah2017} and automating the classification of satellite cloud imagery with high accuracy and efficiency~\cite{bai2021lscidmr}. 

Building upon these developments, the field is now moving toward generative AI, which includes Foundation Models and Large Language Models (LLMs)~\cite{sarhaddi2025llms}. These models not only learn from massive multimodal datasets but can also generate new data, assist in hypothesis generation, and simulate plausible environmental scenarios. Generative AI holds particular promise for atmospheric sciences in tasks such as data synthesis, knowledge extraction from heterogeneous sources, and support for decision-making under uncertainty.
Figure~\ref{fig:AIgroups} illustrates the hierarchical relationship among various AI domains and highlights some widely used techniques, including ML and DL approaches relevant to atmospheric sciences. As these technologies continue to evolve, they promise to further enhance our understanding of atmospheric phenomena and improve our ability to respond to environmental challenges.

\subsubsection{Machine Learning}

Machine learning forms the core of most AI applications in atmospheric sciences, enabling the analysis of complex datasets and the extraction of valuable insights, making it essential for advancing research and operational applications. ML can be broadly classified into four categories: supervised learning, unsupervised learning, semi-supervised learning, and reinforcement learning.

\textbf{Supervised learning} relies on labeled data, where models are trained using input-output pairs to make predictions or classifications. The availability of high-quality labeled data is crucial for the success of these models, as it directly affects their performance and accuracy~\cite{aula2022evaluation}. Algorithms such as support vector machines and neural networks have been extensively applied in atmospheric sciences for tasks such as weather forecasting~\cite{zhao2021hourly} and classifying sources influencing air quality indices~\cite{mad2015classifying}. 

\textbf{Unsupervised learning}, in contrast, deals with unlabeled data, identifying hidden structures or patterns within datasets. Methods such as k-means and hierarchical clustering are commonly used in atmospheric science to analyze air pollutants, including identifying pollution sources, investigating long-range pollutant transport pathways, and supporting the development of effective mitigation strategies~\cite{govender2020application}. 

\textbf{Semi-supervised learning} bridges the gap between supervised and unsupervised methods by leveraging a small amount of labeled data along with a larger pool of unlabeled data. This approach is particularly valuable in atmospheric applications where labeled data is scarce or expensive to obtain, such as detecting extreme weather events using multichannel spatiotemporal convolutional neural networks~\cite{racah2017}. 

\textbf{Reinforcement learning}, a goal-oriented learning paradigm where an agent interacts with its environment and learns through feedback in the form of rewards or penalties, has thus far been largely underexplored in atmospheric sciences due to challenges in creating realistic, interactive environments and the computational demands of simulations. However, reinforcement learning holds promise for enabling automated decision-making and actions derived from atmospheric monitoring and modeling, a topic we return to in Section~\ref{sec:roadmap}. 

\begin{figure*}[h]
    \centering   \includegraphics[width=0.99\textwidth]{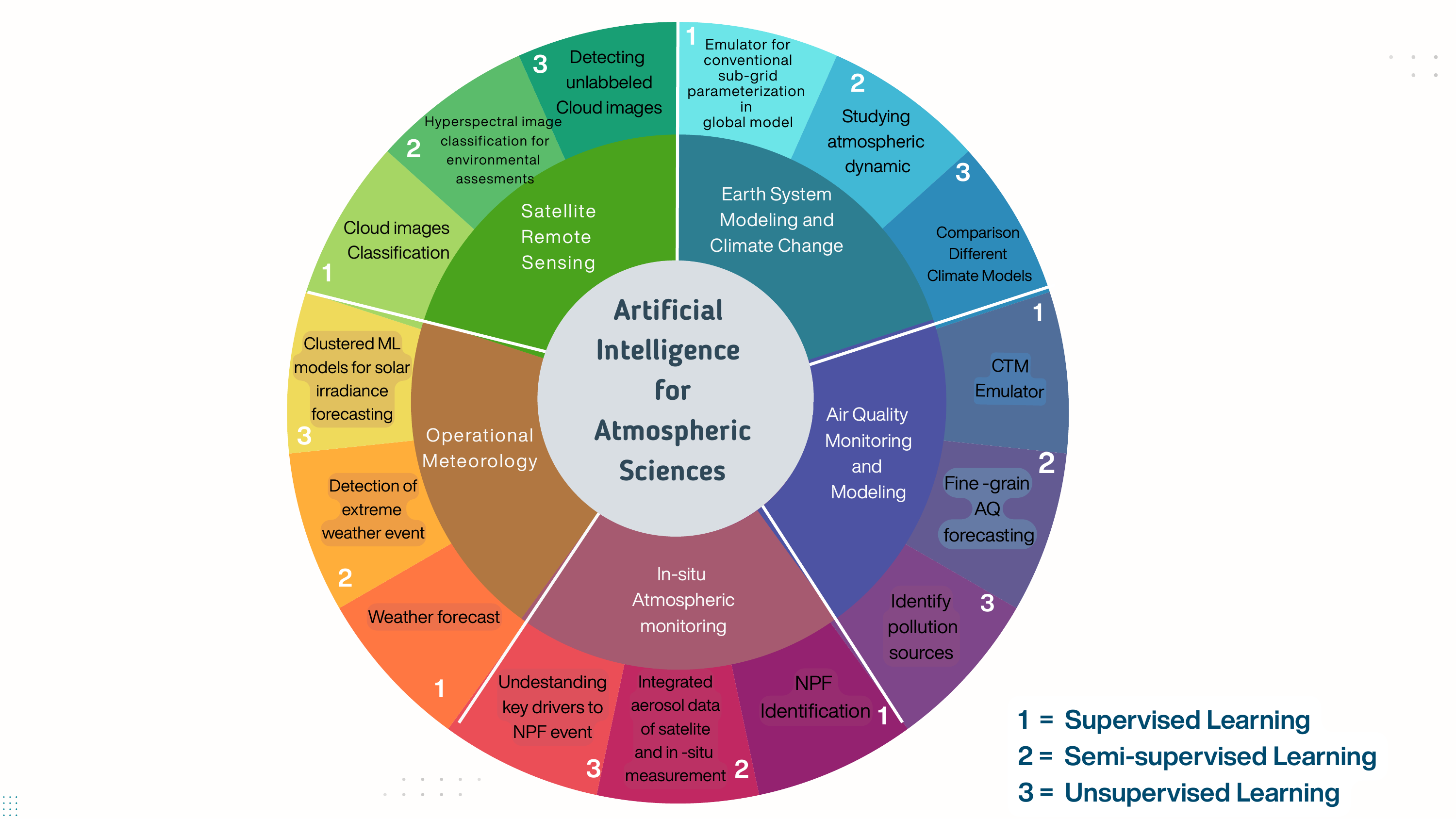}
    \caption{The applications of supervised, semi-supervised and unsupervised learning methods in different fields of atmospheric sciences. 
    }
    \label{fig:ATMbranchesclasses}
\end{figure*}

Figure~\ref{fig:ATMbranchesclasses} illustrates the applications of different learning methods in atmospheric sciences and the following subsections explore their applications across different fields within atmospheric sciences. Understanding these various ML methodologies is crucial for harnessing their potential to address complex challenges in atmospheric research and operational applications, and to identify the future research directions.


\subsection{Air Quality Monitoring and Modeling}

Air quality monitoring and modeling refer to the processes and techniques used to observe, analyze, and predict the concentrations of air pollutants in the atmosphere. 
Common pollutants include particulate matter (\ch{PM}), carbon monoxide (\ch{CO}), nitrogen oxides (\ch{NOx}), sulphur dioxide (\ch{SO2}), ozone (\ch{O3}) and  particle number concentration (\ch{PNC}). These pollutants have significant implications for public health~\cite{shaban2016urban,fung2022improving}, ecosystems~\cite{von2020will}, and climate~\cite{fowler2020chronology}.

Monitoring involves the use of various sensors and instruments, such as those described in the previous section, placed at ground-based stations or integrated into satellite systems to collect real-time data on air pollutants~\cite{zhou2023review}. In contrast, modeling employs mathematical and computational methods to simulate air quality scenarios, identify pollution sources, predict future pollutant levels, and assess the impact of various factors such as weather patterns, urbanization, and industrial activities~\cite{baklanov2020advances}.
The most widely adopted practice for air quality observation have relied on ground-based monitoring stations~\cite{singh2021sensors} and chemical transport models (CTMs)  to simulate the dispersion and chemical transformation of pollutants~\cite{silibello2014application}. However, these methods face challenges, including limited spatial resolution~\cite{zaidan2020intelligent} and high computational costs~\cite{sokhi2021advances}. 

In recent years, the integration of AI into air quality monitoring and modeling has transformed the field.  AI-based models have been used to emulate the outcomes of extensive CTM numerical calculations at significantly lower computational costs~\cite{vlasenko2021simulation}. For instance, deep learning methods have been employed to emulate a typical gas-phase chemistry solver implemented in CTMs, achieving computational efficiency improvements of 10.6 times on one CPU and 85.2 times on one GPU \cite{liu2021emulation}. These substantial gains enable more extensive simulations and faster analyses, highlighting the transformative potential of AI in atmospheric research.

ML and DL techniques have also been employed to predict air quality levels with higher accuracy and efficiency than traditional models can achieve~\cite{liu2021intelligent}. For example, supervised learning algorithms such as support vector machines (SVMs)~\cite{leong2020prediction} and random forests~\cite{zhan2018spatiotemporal} have been used to forecast pollutant concentrations based on historical data and meteorological variables. Similarly, deep learning models such as convolutional neural networks (CNNs)~\cite{chauhan2021air} and long short-term memory (LSTM) networks~\cite{jin2021multivariate} have demonstrated success in capturing complex, non-linear relationships between pollutants and environmental factors. 
Semi-supervised learning has been employed to achieve fine-grained air quality forecasting. For example, self-supervised hierarchical graph neural networks can enhance predictive accuracy even in scenarios with limited labeled data, thereby expanding the applicability of AI-driven models in diverse monitoring environments~\cite{han2022semi}.

The trend to improve air quality forecasting systems has led to the development of new methods that utilize modern observational data in models, including data assimilation and ML methods that combine and fuse information from multiple sources. These sources include low-cost sensor and medium-cost sensor networks, permanent monitoring networks, and UAV-based, aircraft-based, and satellite-based measurements (both in situ and remote sensing) ~\cite{baklanov2020advances}. 
This multi-source data fusion enhances the spatial and temporal resolution of air quality predictions and fosters synergies between atmospheric sciences and computational technologies.
Other AI techniques such as clustering and anomaly detection have also proved useful, e.g, to identify pollution sources~\cite{austin2012framework}, track their dispersion~\cite{huang2021overview}, and even detect unusual pollution events~\cite{borah2024deep}. In urban environments, where industrial activities and traffic are major contributors to air pollution, these AI techniques have been particularly useful in understanding the impact of urbanization and industry on atmospheric composition.

Despite these advancements, several challenges remain in the integration of AI with air quality monitoring. Data scarcity and data quality issues can introduce biases into AI models~\cite{liang2023integrating}. Common concerns include noise and gaps in sensor measurements. The interpretability of AI is another significant concern as many of the models operate as "black boxes" that offer limited information into the underlying physical processes governing air pollution~\cite{tasioulis2023reviewing}  or the limitations of sensor technology~\cite{concas2021low}.  Another challenge is the need for AI models to generalize well across different geographic regions and climatic conditions, as pollutant behavior can vary significantly~\cite{fu2023innovative,aula2022evaluation}.

One of the key future directions for air quality monitoring and modeling is the development of hybrid models that combine the strengths of AI and  CTMs. These hybrid models aim to leverage the data-driven accuracy of AI while maintaining the interpretability and robustness of physical models \cite{xu2021machine}. This approach exemplifies how bridging AI and atmospheric sciences can enhance both fields, leading to a deeper understanding of atmospheric phenomena and improved predictive capabilities. Another important direction is the deployment of Internet of Things (IoT) devices with monitoring capabilities, which is expected to significantly enhance the field by providing near-real-time predictions and actionable insights for decision-makers~\cite{zaidan2022dense}. These efforts are critical for developing more effective environmental monitoring systems and boosting disaster prediction and climate change mitigation efforts. 

\subsection{In-situ Atmospheric Monitoring} 

In-situ atmospheric monitoring involves to the direct, continuous observation and collection of data on various atmospheric variables at ground-based monitoring stations. Unlike traditional air quality monitoring, which primarily targets pollutants in urban areas, in-situ atmospheric monitoring tracks a wider range of atmospheric variables, such as aerosols, greenhouse gases (e.g., \ch{CO2} and \ch{CH4}) and other trace gases \cite{philipona2021networks}. 
The primary goal is to provide high-resolution, real-time data that accurately reflect the atmospheric composition at specific locations, aiding in climate modeling and weather prediction. This detailed data serves as a crucial input for understanding atmospheric processes, validating remote sensing observations, and improving the precision of climate change projections~\cite{bluestein2022atmospheric}.
As an example initiative, the Aerosol, Clouds, and Trace Gases Research Infrastructure (ACTRIS) is a pan-European research infrastructure that produces high-quality data and information on short-lived atmospheric constituents and the processes driving their variability in both natural and controlled atmospheres. ACTRIS stations are typically equipped with particle number counters to monitor atmospheric composition. Figure~\ref{fig:actris} illustrates the locations of several atmospheric research stations that are used to perform extensive measurements of atmospheric concentrations.

Historically, data analysis in atmospheric science has relied on statistical methods. However, the increasing volume and complexity of modern datasets demand advanced AI methods to process and analyze the vast quantities of data ground-based measurement stations generate~\cite{petzold2024opinion}. ML techniques have emerged as valuable tools, capable of automating data analysis, detecting trends, identifying anomalies, and predicting future atmospheric events~\cite{dueben2022challenges}.
For instance, ML and DL models have been successfully applied to study atmospheric new-particle formation (NPF), a key process in generating climatically significant aerosol particles. NPF events, detected through changes in aerosol particle size distributions, traditionally require manual classification of measurement days which is time-intensive and prone to human bias~\cite{kulmala2012measurement, kulmala2013direct, dal2005formation}. Recently, automated ML-based classification of NPF events has significantly reduced manual effort and enabled the rapid creation of event datasets, facilitating further analysis by atmospheric scientists~\cite{zaidan2018predicting, su2022new}.

Unsupervised learning can also be highlight helpful at automating manual analysis processes. For example, mutual information~\cite{marinoni2017unsupervised} has been used to quantify non-linear relationships between various measured variables and NPF events~\cite{zaidan2018exploring, laarne2022exploring}. Methods such as these combined with big data analytics frameworks enable scientists to mine these relationships, uncover associations, and generate new hypotheses that might otherwise remain hidden.
Semi-supervised learning methods can also reduce manual effort, particularly in cases where large volumes of unlabelled data are supplemented by a limited number of labeled samples. For example, semi-supervised learning methods have been used to integrate aerosol data from multiple satellite instruments alongside high-accuracy, ground-based instruments, enhancing the reliability and scalability of aerosol predictions~\cite{djuric2013}.

Despite these advancements, significant challenges remain in applying AI to in-situ atmospheric measurement data. Similar to air quality modeling, the black-box nature of ML models is a major concern, as it complicates the interpretation of their decision-making processes and obscures the underlying atmospheric dynamics. This lack of transparency limits trust and hinders the adoption of AI-driven insights in critical applications. Moreover, current data science methods often overlook the importance of learning cause-and-effect relationships from observations, focusing instead on correlations without considering the physical and generative processes that govern atmospheric systems. Such limitations reduce the generalizability and scientific interpretability of AI-based approaches~\cite{tuia2021toward}.

To address these challenges, future research must prioritize the development of AI models that are both interpretable and trustworthy. Integrating explainable AI (XAI) techniques can provide the necessary transparency in decision-making, enabling researchers to trace model outputs back to their contributing factors~\cite{flora2024machine}. This enhances model credibility and fosters confidence in AI-driven atmospheric predictions. Furthermore, incorporating causal inference methods into AI frameworks is essential to ensure that models learn meaningful cause-and-effect relationships rather than relying solely on statistical correlations~\cite{kukreja2024advancing}. Such advancements are crucial for enhancing our understanding of atmospheric dynamics. In addition, developing interactive AI systems will allow human researchers to query and scrutinize the internal reasoning of models, facilitating collaboration between domain experts and AI-driven analytics~\cite{dueben2022challenges}.

\subsection{Operational Meteorology}



Operational meteorology refers to the application of meteorological science in real-time settings to support decision-making, forecasting, and public safety~\cite{harrison2025assessment}. This involves the use of observational data, numerical weather prediction (NWP) models, and forecasting tools to produce weather forecasts and warnings that are disseminated to governments, industries, and the public. 
Accurate weather forecasting plays a critical role in agriculture, aviation, disaster preparedness, and daily life, making it an essential part of modern society~\cite{chen2024machine}. Standard practice in operational meteorology involves using NWP models that employ mathematical equations to simulate atmospheric processes~\cite{schultz2021can, bi2023accurate}. However, the advent of high-performance computing (HPC), satellite observations, and the Internet of Things, has lead to an explosion in the volume of atmospheric data available for analysis, presenting both challenges and opportunities for the field~\cite{wei2022status}.

AI techniques, particularly ML and DL, have revolutionized weather forecasting by enabling the analysis of vast amounts of data from diverse sources, including ground-based stations, weather radars, satellite remote sensing, and meteorological models~\cite{espeholt2022deep}. AI models excel at handling large datasets, identifying complex patterns, and making data-driven predictions that enhance traditional weather forecasting methods~\cite{schultz2021can}. 
For example, convolutional neural networks (CNNs) have been applied to satellite images to improve cloud classification and storm tracking~\cite{jena2022deep,wang2021tropical}. Similarly, recurrent neural networks (RNNs), especially long short-term memory (LSTM) networks, have been utilized to model temporal dependencies in weather data, leading to more accurate short-term weather predictions~\cite{suleman2022short}.

One of the most transformative applications of AI in operational meteorology is enhancing ensemble weather forecasting, which traditionally involves running multiple simulations with slight variations in initial conditions to account for atmospheric uncertainties~\cite{gronquist2021deep}. AI-driven models, particularly generative adversarial networks (GANs), have been integrated into the post-processing of ensemble forecasts, significantly reducing error margins and increasing the reliability of probabilistic predictions~\cite{bihlo2021generative}. In addition, AI has improved nowcasting, the short-term forecasting of weather over the next few hours, by processing real-time radar and sensor data to deliver accurate predictions on rapid timescales, which is essential for severe weather warnings~\cite{bojinski2023towards}.

Semi-supervised learning techniques, such as a multichannel spatiotemporal CNN architectures, have advanced the detection of extreme weather events such as hurricanes, extra-tropical cyclones, and weather fronts, particularly when available data lacks comprehensive labeling~\cite{racah2017}. Unsupervised learning has also been applied to short-term solar irradiance forecasting. 
Forecasts are generated by using the most suitable models in different clusters that help to improve the accuracy of one-hour-ahead forecasts for global horizontal irradiance (GHI), a key parameter for solar energy management~\cite{feng2018unsupervised}. These AI applications illustrate the growing role of machine learning techniques across varied facets of operational meteorology, from managing uncertainties in ensemble forecasting to making real-time predictions and optimizing solar forecasting.

Despite these advancements, challenges persist in the application of AI to operational meteorology. As before, the limited interpretability of AI models is a critical concern, raising concerns for operational meteorologists who need to trust and understand the models they rely on for decision-making~\cite{yang2024interpretable}. Another challenge stems from the inherent unpredictability of the atmosphere. Specifically, while AI models excel at capturing patterns from historical data, they often struggle to generalize to novel atmospheric conditions, particularly in the face of climate change and extreme weather events~\cite{montillet2024big}. Furthermore, integrating AI with traditional NWP models requires substantial computational resources, making it challenging to balance the strengths of physics-based models with the efficiency of data-driven approaches~\cite{kochkov2024neural}.  

To address these challenges, future research should focus on developing hybrid models that combine the physical insights of NWP with the predictive capabilities of AI. These models aim to leverage the strengths of both approaches, enhancing forecast accuracy while preserving the interpretability and reliability required for operational meteorology~\cite{kochkov2024neural}. AI can also be integrated into specific tasks such as bias correction, downscaling, and data assimilation, ensuring that physics-based constraints remain central to the forecasting process. In addition, advancements in edge computing and IoT technologies could enable real-time data collection and AI-driven analysis, allowing for more precise localized weather predictions with reduced reliance on centralized computational resources \cite{zaidan2024irmaset}. These innovations will facilitate more efficient and scalable meteorological models, bridging the gap between AI and traditional forecasting techniques.  

\subsection{Satellite Remote Sensing}


Remote sensing involves acquiring data about the Earth's surface and atmosphere using sensors mounted on satellites or airborne platforms. These sensors detect electromagnetic radiation that is either reflected or emitted from the Earth, which is subsequently processed into images or datasets that reveal critical environmental parameters~\cite{sabins2020remote}. 
Satellite remote sensing plays a vital role in atmospheric sciences, particularly in addressing global-scale monitoring gaps that cannot be covered by ground-based sensors alone. For example, over oceans, which constitute more than 70\% of the Earth's surface, there is a lack of fixed air quality towers or IoT sensors, making satellite observations indispensable. Remote sensing enables the measurement of atmospheric pollutants such as aerosols, nitrogen dioxide (\ch{NO2}), and ozone (\ch{O3}) over both land and sea, supporting air quality assessments even in remote or data-sparse regions~\cite{fu2022improving}. 
In addition, satellite-based measurements are critical for understanding ocean-atmosphere interactions, including ocean acidification and carbon fluxes. These systems help quantify oceanic uptake of \ch{CO2}, shedding light on the ocean's role as a carbon sink and its influence on global climate dynamics~\cite{shutler2024increasing}.

\begin{figure}[htbp]
\centering
\includegraphics[width=0.475\textwidth]{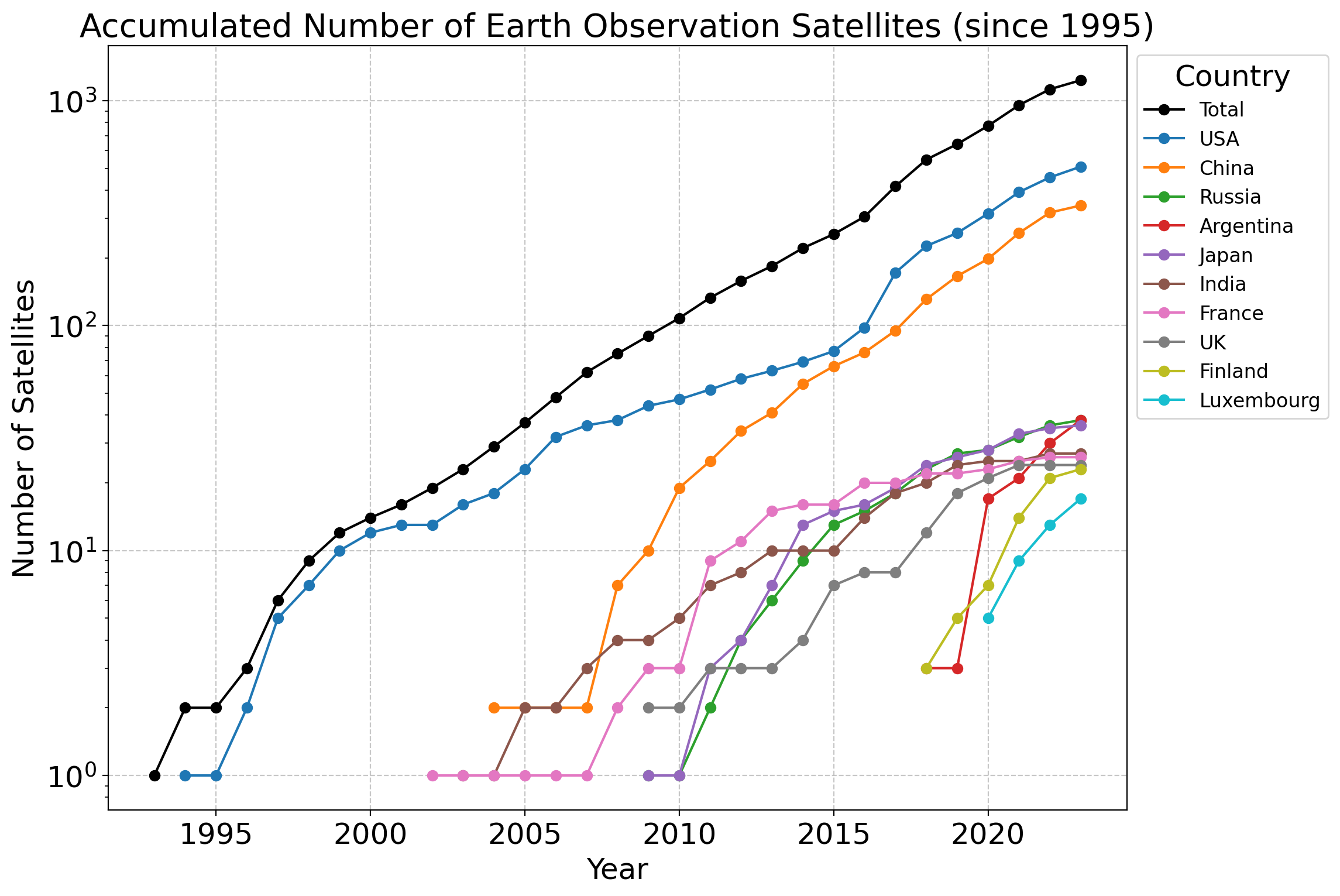}
\caption{Accumulated number of Earth observation satellites (since 1995)}
\label{fig:satellite_remote_sensing}
\end{figure}

Figure~\ref{fig:satellite_remote_sensing} illustrates the accumulated number of Earth observation satellites (since 1995). Over the past three decades, the number of satellites dedicated to Earth observation has surged, significantly enhancing the capability to monitor various aspects of the Earth's atmosphere, oceans, land surfaces, and vegetation. Satellite remote sensing has become indispensable for climate change research, disaster management, agriculture, and weather forecasting~\cite{avtar2020assessing,zhang2022progress}. 

In recent years, AI has played an increasing role in enhancing the analysis of satellite remote sensing data, allowing for more detailed environmental monitoring and assessment~\cite{zhang2022artificial}. By applying AI to extract meaningful features from vast, complex datasets, researchers can improve the spatial and temporal resolution of remote sensing applications, which results in more accurate insights into environmental changes~\cite{han2023survey}. 
For instance, convolutional neural networks (CNNs) are widely used for the automated classification and recognition of satellite cloud images, as demonstrated on large-scale cloud image databases~\cite{bai2021lscidmr}. In tandem, recurrent neural networks (RNNs) and long short-term memory (LSTM) networks analyze temporal sequences within satellite data, enabling better predictions of atmospheric variables and the monitoring of dynamic events like forest fires and deforestation~\cite{persello2022deep,miller2024deep}. 
Semi-supervised learning techniques are utilized for tasks like hyperspectral image (HSI) classification, where limited labeled data restricts traditional training. These techniques enable enhanced feature extraction and classification for detailed environmental assessments~\cite{cao2020sdfl,wang2022self}. Unsupervised learning methods are increasingly critical in remote sensing, particularly in point cloud analysis, where the absence of labeled data limits traditional approaches. For example, unsupervised point cloud detection has emerged as a key technique, enabling generalized representations from unlabeled data that effectively support diverse applications in environmental monitoring~\cite{guo2021unsupervised}.

One key advancement enabled by AI is the improvement in the spatial and temporal resolution of satellite-derived products. Traditional remote sensing systems often face limitations in their ability to provide high-resolution data due to trade-offs between spatial coverage and frequency of observations. AI techniques, such as super-resolution models, have been developed to enhance the resolution of satellite images, allowing for more detailed monitoring of small-scale environmental changes~\cite{wang2022comprehensive,sdraka2022deep}. Furthermore, AI-based algorithms have been employed to fuse data from multiple satellites and sensors, creating comprehensive datasets that offer more accurate and holistic insights into the Earth's atmospheric and environmental systems~\cite{Himeur2022,Zhao2024}.

Despite these advancements, several challenges remain in the application of AI to satellite remote sensing. The primary challenge is the vast volume and diversity of data generated by multiple satellites, which requires powerful computational resources and efficient algorithms to process~\cite{xu2022cloud}. Furthermore, AI models used for remote sensing data analysis often require large, annotated training datasets, which are not always readily available~\cite{schmitt2023there}. Finally, as with the other fields, also remote sensing would require for explainable models that allow interpreting the physical meaning of their outputs~\cite{wang2024trustworthy}.

To overcome these challenges, future research should focus on the development of explainable AI-based models and hybrid models that combine the strengths of both AI and traditional remote sensing techniques~\cite{zhang2022artificial}. In addition, efforts to create more comprehensive, annotated datasets will be crucial for improving the performance of AI models in satellite data analysis~\cite{bai2021lscidmr,schmitt2023there}. As computational power continues to grow and more advanced AI algorithms are developed, the integration of AI with satellite remote sensing will likely yield even greater improvements in environmental monitoring and climate science.

\begin{table*}[!htbp]
  \centering
    \caption{Summary of AI applications in atmospheric sciences}
    \label{tab:ai_atmospheric_summary}
    \begin{tabularx}{\textwidth}{p{0.15\linewidth}XXX}
    \toprule
       \textbf{Field} & \textbf{State-of-the-Art} & \textbf{Challenges} & \textbf{Possible Solutions}  \\
    \midrule    
\textbf{Air Quality Monitoring and Modeling} & 
AI enhances air quality predictions, source identification, and pollutant dispersion modeling. ML and DL methods improve computational efficiency over chemical transport models (CTMs) and detect pollution anomalies. &
Data scarcity, model interpretability, and generalization across regions with diverse pollution dynamics.  &
Hybrid AI-physics models, expanded sensor networks, and real-time IoT-based monitoring for improved accuracy. \\
\hline
\textbf{In-situ Atmospheric Monitoring} &
AI automates data analysis, trend detection, and anomaly identification in ground-based observations, improving real-time monitoring. &
ML models often act as black boxes, limiting transparency. Causal relationships in atmospheric data remain underexplored. &
Explainable AI (XAI) for model interpretability, causal inference techniques, and interactive AI-driven analytics for enhanced scientific collaboration. \\
\hline
\textbf{Operational Meteorology} & 
AI-driven models refine weather forecasts using ML and DL, integrating multi-source data for extreme weather event prediction and downscaling. &
Lack of interpretability in AI forecasts, generalization issues under novel conditions, and high computational costs. &
Hybrid AI-NWP models, physics-constrained AI, and edge computing for real-time weather analysis and enhanced extreme event forecasting. \\
\hline
\textbf{Satellite Remote Sensing} &
AI extracts features from vast satellite datasets, enhancing image classification, hyperspectral analysis, and multi-modal fusion for environmental monitoring. &
Processing large datasets, limited labeled training data, and AI model transparency for physical interpretation. &
Explainable AI, hybrid AI-remote sensing approaches, and improved annotated datasets for enhanced AI training. \\
\hline
\textbf{Earth System Modeling and Climate Change} &
DL improves climate simulations, replaces conventional parameterizations, and enhances subgrid process modeling. Unsupervised learning refines climate variability analysis. &
Interpretability of DL models, maintaining physical consistency, handling uncertainty, and computational costs. &
Physics-informed neural networks (PINNs), scalable DL models on HPC/cloud, and probabilistic AI methods for uncertainty quantification. \\
\bottomrule
    \end{tabularx}
\end{table*}

\subsection{Earth System Modeling and Climate Change}

Earth system modeling (ESM) is a multidisciplinary approach to simulating the complex interactions between the Earth's atmosphere, oceans, land, and biosphere. These models integrate various physical, chemical, and biological processes to study the behavior of the Earth as a single, interconnected system~\cite{Bonan2018}. One of the primary objectives of ESM is to improve our understanding of climate change by simulating future climate scenarios based on different greenhouse gas emission trajectories~\cite{gettelman2022future}. Climate change, driven by human activities such as industrialization and deforestation, has led to significant changes in the Earth's climate, including rising temperatures, shifting weather patterns, and an increasing frequency of extreme weather events~\cite{Trenberth2015}. ESMs are critical for predicting how these changes will evolve over the coming decades and centuries, providing essential information for policymakers and scientists working on climate mitigation and adaptation strategies~\cite{Hewitt2021}.

DL models have recently emerged as powerful tools for improving the accuracy and efficiency of Earth system modeling, particularly in climate prediction and the study of climate change~\cite{Irrgang2021}. Traditional ESMs are computationally intensive and rely heavily on physics-based equations to simulate the Earth's climate. Although these models provide detailed insights into physical processes, they often struggle with high computational costs and long simulation times. In contrast, DL models, such as CNNs and RNNs, have shown promise in capturing complex, non-linear patterns in climate data, enabling faster and more efficient simulations~\cite{reichstein2019deep}.
For instance, DL  techniques have been applied to represent subgrid atmospheric processes in climate models, learning directly from multiscale models that explicitly resolve convection. These trained DL models can replace conventional subgrid parameterizations in global general circulation models, allowing them to interact dynamically with resolved atmospheric processes and surface-flux schemes. This approach significantly enhances the representation of complex, nonlinear subgrid processes, particularly cloud dynamics, which have historically been a significant source of uncertainty in climate models~\cite{rasp2018deep}. Another example of DL in ESMs is its use in forecasting the El Niño/Southern Oscillation (ENSO); a DL-based statistical model has shown high accuracy in predicting ENSO patterns, which is crucial for anticipating extreme regional climate events and their ecosystem impacts~\cite{ham2019deep}.

To address the common challenge of limited labeled data, semi-supervised learning (SSL) methods are also being developed for diverse atmospheric applications. SSL is particularly valuable in studying atmospheric dynamics, where large-scale labeled datasets are sparse, enabling models to leverage both labeled and unlabeled data to enhance predictive performance across atmospheric variables~\cite{HoffmannLessig2023}.
Unsupervised learning methods, such as variational autoencoders (VAEs), offer further advancements by performing dimensionality reduction and density estimation on climate data. VAEs automatically derive low-dimensional, physically meaningful representations of complex datasets from various climate models, uncovering new notions of similarity and variability in atmospheric patterns~\cite{mooers2023comparing}.

In addition to improving predictive accuracy, interpretability and understanding are essential for the effective use of DL in Earth system models. Interpretability remains a challenge for deep neural networks, as they are often not self-explanatory and may struggle to identify causal relationships from observational data. Even modern Earth system models suffer from limited traceability to their assumptions, which reduces their interpretability. Ensuring physical consistency in predictions is another important challenge. Specifically, while deep learning models fit data well, they may produce physically implausible outcomes due to observational biases or extrapolation. Integrating domain knowledge and physical laws into models can address this issue. Furthermore, DL must effectively handle complex, noisy, and high-dimensional data, necessitating new network architectures that capture both local and long-range relationships. Managing uncertainty will require integrating Bayesian or probabilistic methods. In addition, many DL applications face the challenge of limited labeled data, highlighting the need for methods that leverage unlabeled data through unsupervised or semi-supervised learning. Finally, the computational demands of geoscience problems are immense, as demonstrated by platforms like Google’s Earth Engine, which can help manage these challenges and may lead to further deep learning applications~\cite{reichstein2019deep}.

In terms of advancements, ongoing research aims to improve the scalability of DL models for global-scale climate simulations, leveraging high-performance computing (HPC) and cloud-based platforms~\cite{li2020geospatial,mazzoglio2022exploitation}. Furthermore, efforts are being made to develop more interpretable AI models, such as physics-informed neural networks (PINNs), which incorporate physical laws into DL architectures to enhance both accuracy and transparency~\cite{de2021assessing,meray2024physics}.

\subsection{Other Environmental Challenges and Atmospheric Science Research}


Atmospheric science intersects with various environmental disciplines, highlighting the broader impact of AI applications beyond traditional meteorological studies. AI techniques developed for atmospheric research can be extended to address other environmental challenges, contributing to a more comprehensive understanding of Earth system processes.  For example, the coupled interactions between the atmosphere, ocean, and land surface form the backbone of Earth system dynamics, influencing climate variability, weather extremes, and long-term environmental trends~\cite{bonan2018climate}. 
AI is increasingly used to capture these cross-domain processes by analyzing multivariate datasets from satellites, in-situ sensors, and reanalysis products. 
For example, ML models help reveal how oceanic conditions such as sea surface temperature and salinity influence atmospheric circulation, while also assessing how land features like vegetation cover, topography, and soil moisture feedback into weather systems and carbon exchange processes~\cite{reichstein2019deep}. These integrative approaches support the development of more robust climate predictions and Earth system models by incorporating interactions that were traditionally studied in isolation.

AI is also revolutionizing extreme weather event prediction by improving the accuracy and timeliness of forecasts for hurricanes, tornadoes, and heatwaves~\cite{lamsal2020artificial}. By processing large-scale datasets from satellites and ground stations, AI-driven models enhance disaster preparedness and response efforts, enabling better risk assessment and mitigation strategies~\cite{albahri2024systematic}.  

The hydrological cycle, which is deeply connected to atmospheric processes, also benefits from AI applications. Machine learning techniques are being utilized to predict rainfall patterns, river flows, and drought conditions, enabling more effective water resource management. Such advancements are crucial for agriculture, urban planning, and ecosystem sustainability~\cite{niu2021evaluating,chang2023artificial}.  
In agricultural meteorology, AI-driven models are used to predict weather patterns and assess their impact on agricultural production. These models help optimize crop yield forecasting and food security planning by analyzing shifts in temperature, precipitation, and other meteorological factors that influence agriculture~\cite{sharma2020machine}.  

Computational atmospheric chemistry is another area where AI plays a transformative role. ML and DL algorithms are increasingly used to simulate complex chemical interactions in the atmosphere, predict pollutant transformations, and enhance air quality models. These AI-driven advancements provide valuable insights into the interplay between atmospheric chemistry, climate change, and air pollution control strategies~\cite{leonardi2020particle}.

\subsection{Summary}

The discussions in the preceding subsections are synthesized in Table~\ref{tab:ai_atmospheric_summary}, which provides a concise overview of current state-of-the-art, challenges, and possible solutions for AI in atmospheric sciences. 
While possible solutions are limited for short future solutions. In the next section, this paper will discuss wider and futuristic aspects of AI-driven solutions for atmospheric sciences in terms of a roadmap. The proposed roadmap serves as generic futuristic long-term solutions which will shape better future for atmospheric sciences and their respective real-world applications.

\section{Roadmap of AI-Driven solutions for Atmospheric Sciences} \label{sec:roadmap}

The integration of AI into atmospheric and environmental sciences presents unprecedented opportunities to enhance research, monitoring, and predictive capabilities. As the demand for real-time, high-accuracy data continues to grow, future research must concentrate on key directions that not only leverage AI-driven solutions but also contribute to the advancement of AI itself.

The complexity of atmospheric phenomena offers rich datasets that can significantly enhance AI research. Simultaneously, improving the quality and scale of these data is vital for a better understanding of atmospheric processes. One essential avenue for exploration is \textit{optimizing sensing infrastructure for future atmospheric systems}. Advances in signal processing, anomaly detection, and self-calibrating sensors can improve operational efficiency of sensing infrastructure while producing high-quality data that serves as valuable training sets for machine learning models. These innovations enable large-scale deployments, from low-cost sensors to sophisticated instrumentation at research stations, fostering a more nuanced understanding of atmospheric dynamics while also providing a wealth of data to challenge AI algorithms.

As the volume and complexity of atmospheric data increase, \textit{next-generation computing platforms for atmospheric data processing} will play a crucial role. By combining high-performance computing, cloud computing, and edge-fog platforms, we can process massive datasets with minimal latency, which is vital for both atmospheric research and AI development. This capability allows researchers to conduct real-time simulations and analyses, pushing the boundaries of AI in processing and interpreting complex systems, thus driving innovation in AI methodologies. At the same time, faster and more reactive data processing enhances atmospheric fields, supporting applications from real-time weather forecasting to disaster response and long-term climate modeling. 

Emerging directions in AI, referred to as \textit{advanced AI methodologies, such as foundation models and explainable AI}, can also benefit from atmospheric sciences. Large-scale foundation models trained on multi-modal datasets derived from atmospheric research can uncover intricate patterns and relationships that individual modalities may miss, improving our understanding of atmospheric phenomena. In parallel, the diverse and dynamic nature of atmospheric data presents unique challenges that can inspire the development of new AI algorithms and techniques. Moreover, incorporating explainable AI approaches ensures transparency and trust, particularly in high-stakes applications such as climate predictions and public health policy.

To effectively leverage these advancements, it is crucial to also support the end-users of the information through an AI-driven transition, referred to as \textit{from data to action: analytics, visualization, and automated feedback for environmental decision-making}. Atmospheric sciences benefit from AI-driven insights that translate into actionable outputs for decision-makers, scientists, and the public. Conversely, AI research gains from the practical scenarios and case studies provided by atmospheric sciences, illustrating the real-world implications of AI insights. For instance, automated feedback loops can enhance operational efficiency in optimizing indoor air quality or guiding policy interventions to mitigate harmful emissions, while also refining AI models based on user interactions and outcomes.

Finally, the increase in data collection raises significant legislative concerns about data management, usage, and collection practices. \textit{Data governance, privacy, and collaboration} highlight the importance of ethical AI deployment and robust governance frameworks, which are particularly relevant in atmospheric sciences where data transparency is crucial. By ensuring data privacy, compliance, and equitable access to resources, interdisciplinary collaboration among scientists, policymakers, and industry leaders can be fostered, driving mutual benefits for both fields.

\begin{figure*}[htbp]
    \centering
    \includegraphics[width=0.9\linewidth]{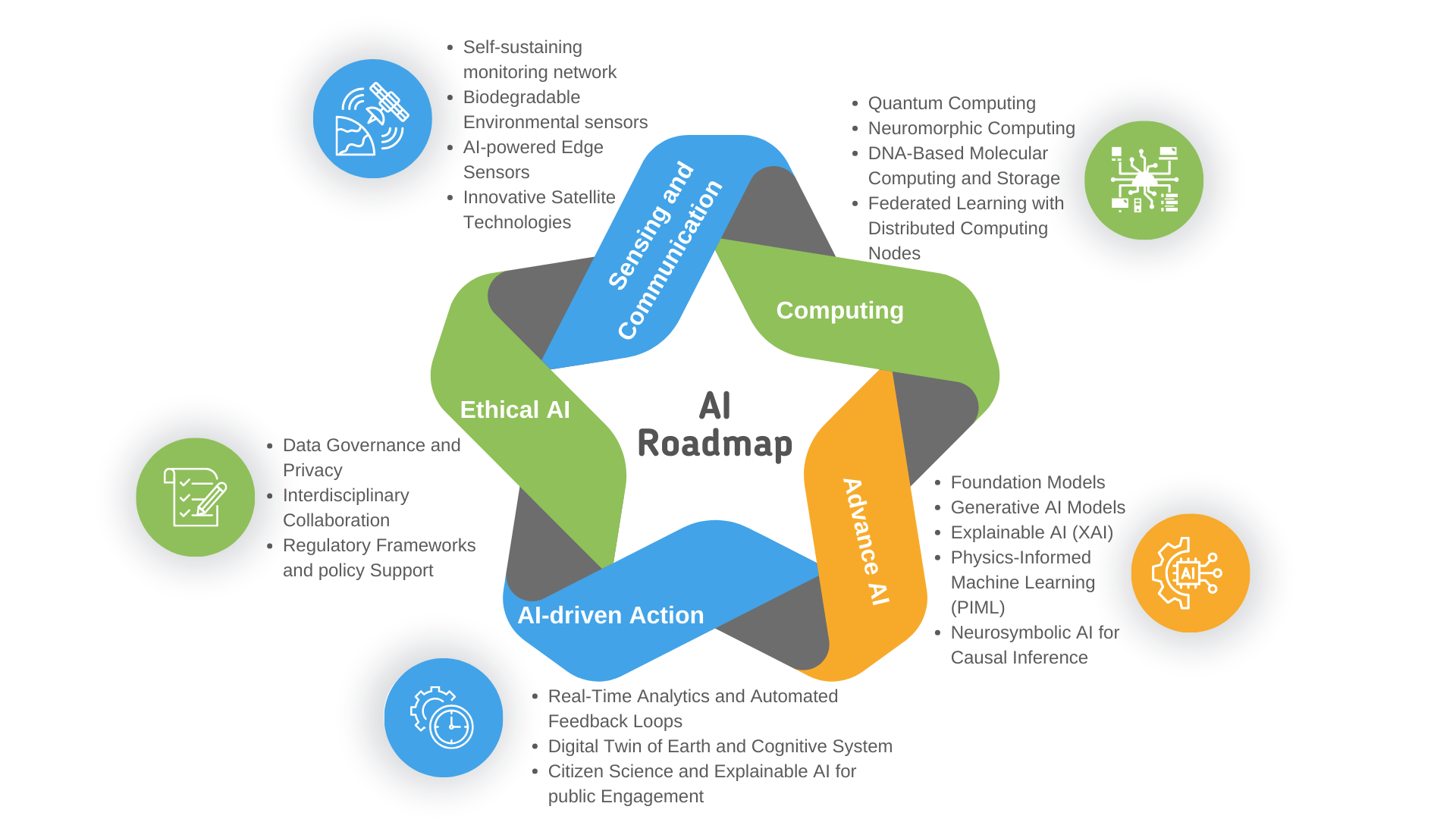}
    \caption{Roadmap of AI-driven solutions and research potentials for atmospheric sciences}
    \label{fig:roadmap}
\end{figure*}

Figure~\ref{fig:roadmap} summarizes these interconnected trends, illustrating how atmospheric sciences not only benefit from AI integration but also provide fertile ground for the advancement of AI research itself. This reciprocal relationship holds the potential to fuel the next generation of innovations in both atmospheric and AI sciences.

\subsection{Sensing and Communication Infrastructure}

The future of atmospheric monitoring demands groundbreaking innovations in the sensing and communication infrastructure to address the challenges of real-time, scalable, and sustainable data collection. 

\subsubsection{Self-sustaining monitoring networks}

Energy inadequacy is a significant barrier to the long-term operation and deployment of atmospheric sensing systems, particularly in remote, inaccessible, or harsh environments. To address this challenge, there has been increasing interest in self-powered sensors capable of harvesting energy from their surroundings~\cite{javaid2023self}. These sensors can scavenge energy from environmental sources, such as sunlight, radio frequency (RF) electromagnetic waves, thermal gradients, and mechanical energy, enabling sustainable, maintenance-free operations over extended periods~\cite{ahn2024innovations}.

While solar and RF harvesting have long been investigated in the context of environmental monitoring~\cite{ferdous2016renewable}, significant challenges remain in transforming these technologies into self-sustained monitoring networks. Specifically, the quantities of energy harvested tend to be relatively low, often sufficient only for simple operations such as gathering individual readings and transmitting them, rather than for storing and processing data. Additionally, sampling certain atmospheric compounds requires energy-intensive techniques that exceed the capabilities of existing harvesting methods~\cite{concas2021low}.

Beyond solar and RF harvesting, recent advancements in ocean wave energy conversion present promising possibilities for powering ocean-based IoT devices. These are crucial for continuous monitoring of marine environments and understanding atmospheric-ocean interactions that significantly influence climate systems and weather predictions~\cite{de2023investigation}. By providing high-resolution, real-time data on oceanic conditions, these self-powered sensors enhance the quality of atmospheric models and improve forecasting accuracy, supplying valuable data for machine learning algorithms in AI applications.

Despite these innovations, substantial challenges persist. Optimizing energy efficiency in sensors and processing systems is critical to complementing energy harvesting, while issues related to sensor size, cost, robustness, long-term stability, and seamless integration with AI-driven data processing systems also need to be addressed~\cite{javaid2023self}. Advancing self-powered sensor technology requires a multidisciplinary approach to improve the durability, adaptability, and energy efficiency of these devices. Continuous innovation in this field is vital for developing resilient and reliable sensing networks capable of meeting the demands of next-generation atmospheric monitoring systems. By integrating AI methodologies and advanced data analytics, researchers can gain deeper insights into atmospheric phenomena, ultimately leading to more effective environmental management and policy-making.

\subsubsection{Biodegradable environmental sensors}

As sensor networks continue to grow, the issue of electronic waste is becoming increasingly critical and necessitating sustainable solutions for the disposal of electronic devices~\cite{ngoy2025supporting,fumeaux2023printed}. This issue is particularly critical for atmospheric and environmental monitoring as deployments scale up. One promising approach is the development of biodegradable sensors made from environmentally friendly materials, which offer a sustainable monitoring solution, especially in ecosystems vulnerable to pollution. These sensors are designed for short-term deployments and naturally decompose after their operational lifespan, significantly reducing their environmental impact. Recent advancements have demonstrated the potential of biodegradable and renewable materials, such as antennas made from sustainable substrates for environmental sensing applications~\cite{zahedi2024biodegradable}. For instance, chemical-based wearable sensors have shown promise in detecting environmental pollutants, but challenges remain in advancing these systems into next-generation environmental sensors that can be widely deployed.

From a materials perspective, a deeper understanding of the properties of nanomaterials is crucial for achieving biocompatibility, self-healing capabilities, and high performance at low costs. In addition, hybrid energy devices with low form factors are desirable for powering these sensors, as they should utilize multiple types of energy storage and harness energy from diverse fuel sources to enable continuous operation. Ensuring high-speed, low-power real-time data communication is also vital, requiring the integration of advanced wireless technologies, such as millimeter-wave frequency-based or optical networks. Furthermore, biosynthesis-based green approaches can be employed for the on-site remediation of environmental contaminants introduced by nanomaterials~\cite{mamun2020recent}. Addressing these challenges and advancing biodegradable sensor technologies paves the way for greener atmospheric monitoring systems that reduce the environmental footprint of sensor networks and enhance their compatibility with fragile ecosystems, supporting sustainable environmental management.

\subsubsection{AI-powered edge sensors}

Integrating lightweight AI models into edge devices enables real-time data processing directly at the source. These sensors can execute critical functions such as signal processing, anomaly detection, and calibration locally. This reduces the need for extensive data transmission to centralized systems and facilitates immediate actions, which is particularly relevant for domains such as disaster prevention or mitigation~\cite{fouda2022lightweight,patel2024trustable}. 
The embedded algorithms on edge sensors are adept at handling complex tasks, including managing noisy data through real-time signal processing, implementing advanced anomaly detection systems, and utilizing machine learning classifiers to address and mitigate sensor drift. Self-calibration techniques further enhance measurement accuracy and extend sensor operational lifespans by automatically correcting sensor drift~\cite{zaidan2023intelligent}.

From a technical standpoint, edge devices face constraints such as limited computational and storage resources, particularly when deploying DL models for real-time data stream processing with low latency. Environmental challenges, including minimizing pollution hazards, traffic waste, resource consumption, and energy usage, add further complexity. Furthermore, cost-related concerns such as high initial capital investments, ongoing operational expenses, and the need to develop viable business models for integrating these technologies into atmospheric monitoring systems must also be addressed~\cite{bibri2024smarter}. Finally, ensuring interoperability among different devices and systems is vital to maximize the effectiveness of edge deployments.

\subsubsection{Innovative Satellite Technologies}

CubeSats, a class of nanosatellites, have emerged as a cost-effective alternative to conventional satellites. These small, modular satellites are constructed using commercial off-the-shelf components, making them both affordable and adaptable for a wide range of applications~\cite{abulgasem2021antenna}. Their scalability and flexibility have positioned CubeSats as a transformative tool for atmospheric monitoring, enabling researchers to tackle challenges that traditional satellite systems often struggle to meet. 

One key advantage of CubeSats is their ability to communicate effectively with one another and with ground stations, facilitating coordinated operations essential for many fields of atmospheric sciences. For example, tasks such as weather prediction, climate change monitoring, and disaster management benefit from coordinated sampling~\cite{saeed2020cubesat}. The deployment of CubeSat constellations offers a scalable solution for high-resolution atmospheric observations, effectively filling gaps left by traditional systems. These constellations provide near real-time insights into localized weather events~\cite{bluestein2022atmospheric} and urban air quality information~\cite{baddock2021understanding}, thereby enhancing decision-making capabilities in environmental management.

Despite their potential, CubeSats face significant technical and hardware-related challenges. Persistent issues include power control, miniaturization, and configuration~\cite{Gregorio2018}. Their small size limits the integration of advanced capabilities, such as next-generation wireless communication systems and sophisticated AI processing capabilities~\cite{saeed2020cubesat}. In addition, the compact design may restrict payload capacity, hindering the deployment of advanced sensors necessary for comprehensive atmospheric analysis. 

Addressing these challenges is crucial for enhancing the performance and reliability of CubeSats, improving the quality of data they deliver, and ensuring they effectively support atmospheric sciences and environmental monitoring. Innovations in power management systems, miniaturization techniques, and data processing methodologies can enable CubeSats to provide higher-resolution, real-time atmospheric data, thereby fueling demand for advancements in AI processing techniques. By overcoming these obstacles, CubeSats can solidify their role as indispensable tools in environmental monitoring and enhanced decision-making frameworks.

\subsection{Computing Infrastructure }

Advancing atmospheric sciences requires robust and scalable computing infrastructure capable of handling vast datasets and complex simulations. As models grow in complexity and the demand for real-time analysis and forecasting increases, innovations in computing ranging from high-performance computing (HPC) to emerging paradigms like neuromorphic and quantum computing will be essential to drive efficiency, reduce energy consumption, and unlock new capabilities for environmental monitoring and predictive modeling.

\subsubsection{Quantum Computing} 


Quantum computing harnesses the principles of superposition and entanglement to perform computations at speeds unattainable by classical computers~\cite{sood2023quantum}. This capability enables quantum computers to address problems currently deemed intractable, even for the fastest classical systems, offering significant potential for advancements in atmospheric sciences. The unique architecture of quantum computing allows for specialized algorithms that can enhance climate modeling and weather prediction~\cite{corcoles2019challenges}.

Recent advancements have demonstrated the successful application of quantum computers in climate modeling and weather prediction~\cite{ashwani2024quantum}. For instance, numerical weather prediction (NWP) executed on quantum platforms has shown promise in enhancing computational speed and improving forecast accuracy~\cite{frolov2017can}. Furthermore, quantum machine learning techniques can refine weather forecasting models, yielding more precise predictions and more effective disaster mitigation strategies~\cite{suhas2023quantum,patil2024utilizing}.

The emergence of Quantum Computing as a Service (QCaaS) within cloud platforms represents a transformative solution for atmospheric sciences, providing on-demand access to quantum resources without necessitating dedicated infrastructure~\cite{nguyen2024qfaas}. This hybrid approach facilitates the seamless integration of classical HPC approaches with quantum systems, enabling the efficient execution of complex and computationally intensive atmospheric algorithms. By synergizing the strengths of both classical and quantum computing, researchers can unlock new capabilities for data analysis and predictive modeling, advancing our understanding and management of atmospheric phenomena. Moreover, the multi-modal nature of atmospheric datasets and the intricate patterns they encompass can stimulate research in quantum-enabled AI and machine learning algorithms, opening avenues for developing innovative quantum algorithms. 

\subsubsection{Neuromorphic Computing}  

Modern computing systems consume vast amounts of energy, rendering them unsustainable for the complex and data-intensive tasks necessary for atmospheric modeling and environmental monitoring. While not all computationally demanding processes require AI or DL, the widespread adoption of DL exacerbates energy concerns. For instance, training a single DL model can consume up to \num{656,347} kilowatt-hours of energy, resulting in approximately \num{626,155} pounds of \ch{CO2} emissions, equivalent to the total lifetime carbon footprint of five automobiles~\cite{luo2023achieving}. Neuromorphic computing has emerged as a promising solution to improve the sustainability of AI by developing systems inspired by the structural and functional characteristics of the human brain. This field aims to overcome the limitations of conventional computing by replicating the parallelism, fault tolerance, and energy efficiency observed in biological neural networks. Neuromorphic systems integrate hardware and software architectures that mimic the behavior of neurons and synapses, enabling more efficient processing of large datasets with significantly lower energy consumption~\cite{Li2024}. 

While neuromorphic computing has demonstrated improvements in areas such as event-based data processing, adaptive control, constrained optimization, sparse feature regression, and graph search~\cite{davies2021advancing}, its applications in atmospheric sciences remain largely unexplored. Neuromorphic computing holds significant promise for advancing AI-driven atmospheric modeling and monitoring systems by facilitating the training of large-scale AI models with reduced energy demands. Furthermore, the integration of neuromorphic computing has significant promise to improve the sustainability of complex AI models. For example, foundation models such as Microsoft's Aurora~\cite{bodnar2024aurora} and Google's GenCast have demonstrated substantial benefits for weather forecasting and climate modeling~\cite{price2024probabilistic}. Neuromorphic computing is also a promising solution for low-power, smart edge processing in extreme environments. This capability is particularly advantageous for on-board AI processing in satellites, where energy efficiency and real-time data analysis are critical~\cite{vineyard2019resurgence}. By enabling AI-powered edge computing directly on satellites, neuromorphic computing reduces the necessity for continuous data transmission to ground stations, thereby enhancing efficiency~\cite{garcia2024advancements}.

\subsubsection{DNA-based Molecular Computing and Storage}  

Significant advances in biotechnology have enabled efficient manipulation of deoxyribonucleic acid (DNA), making it possible to harness DNA as a programmable biological substrate. DNA computing is a recently emerged field that builds on this premise, leveraging DNA for computing, data storage, and communication~\cite{liu2021dna}. DNA operates with exceptional energy efficiency and possesses remarkable longevity and information density, making it a highly attractive medium for a variety of applications~\cite{liu2021dna}. At the same time, the versatility of DNA nanotechnology facilitates the design of programmable DNA-based nanostructures and microstructures capable of performing tasks ranging from molecular computing to data storage~\cite{yang2024dna}.

Despite its significant potential, the application of DNA computing in atmospheric sciences remains limited. One notable exception is the use of spectral encoding and matching algorithms inspired by DNA computing to classify spectral signatures in hyperspectral remote sensing data~\cite{jiao2012artificial}. This approach demonstrates the feasibility of DNA-based techniques in addressing complex environmental data analysis tasks. However, several challenges hinder the broader adoption of DNA computing. Developing complex DNA-based devices that can function within live cells and regulate their behavior is a significant obstacle. Moreover, as the number of DNA strands required for computation grows exponentially with problem complexity, improving the efficiency of DNA synthesis remains a critical area of ongoing research~\cite{yang2024dna}. Although DNA computing is still evolving, it holds immense promise for addressing key challenges in atmospheric sciences, particularly in areas requiring large-scale data storage, energy-efficient processing, and cost-effective solutions. As advancements continue, DNA-based computing could become a transformative tool for managing the vast datasets associated with atmospheric monitoring and modeling.

\subsubsection{Federated Learning with Distributed Computing Nodes}  

Federated learning (FL) enables the training of AI models across decentralized data sources without transferring raw data, preserving privacy and improving computational efficiency~\cite{beltran2023decentralized}. FL can be highly valuable also in atmospheric sciences as it enables localized data processing without compromising security or data ownership. For instance, FL can be employed to analyze crowd-sensed weather data from networks of low-cost, widely distributed weather stations, facilitating precise local forecasts while safeguarding data privacy~\cite{de2024federated}. FL has also shown promise in source term estimation, where deep neural networks identify unknown gas leakage sources in urban environments~\cite{xu2021federated}. Another example is its use in remote sensing where FL can improve privacy, scalability, and classification accuracy~\cite{moreno2024federated}.  

Beyond its privacy benefits, FL can significantly reduce carbon emissions by decentralizing model training, leading to lower overall energy  consumption~\cite{qiu2020can}. FL also holds promise for generating global models that aggregate data from atmospheric monitoring stations, satellites, and edge sensors. Besides preserving data sovereignty, this approach enhances scalability and lowers the environmental impact of AI-driven atmospheric research. However, atmospheric data also pose challenges for FL. For example, atmospheric data are often characterized by strong temporal correlations and are prone to distributional shifts~\cite{aula2022evaluation}, both of which complicate the use of conventional FL techniques. Another key issue is how to manage stragglers, nodes whose training contributions are delayed, without negatively impacting the final model. These challenges are currently active areas of research~\cite{ye2023heterogeneous}, and access to atmospheric datasets can significantly contribute to advancing this research.

\subsection{Advanced AI Methodologies}

The next generation of AI in atmospheric sciences is expected to be driven by advancements in sophisticated models that can handle multimodal datasets and operate across different spatial and temporal scales. This has the potential to reveal phenomena that link different environmental systems and deepen our understanding of atmospheric processes.

\subsubsection{Foundation Models and LLMS} 

Foundation models represent a transformative leap in AI by enabling generalization across diverse tasks and domains through the utilization of vast training datasets. In atmospheric sciences, these models, trained on extensive repositories of climate, weather, and remote sensing data, hold significant potential to enhance tasks such as weather forecasting, environmental monitoring, and anomaly detection~\cite{li2024new}. For instance, developing universal remote sensing foundation models that leverage millions of spectral images can improve applications including multi-label scene classification, semantic segmentation, and land-use change detection~\cite{hong2024spectralgpt}. A notable example is the Aurora foundation model, developed by Microsoft, which has been trained on over a million hours of weather and climate data to produce high-fidelity operational forecasts across diverse atmospheric variables. Aurora has demonstrated exceptional performance, particularly in scenarios characterized by limited training data and extreme weather events.

Large language models (LLMs)~\cite{sarhaddi2025llms}, a significant subset of foundation models, are also gaining traction in atmospheric and geoscience research. These models excel at interpreting complex textual data, generating coherent and contextually relevant responses, and facilitating interdisciplinary collaboration. An example is the GeoGalactica model ~\cite{lin2023geogalactica}, which extends conventional LLM capabilities by further pre-training on geoscience texts and fine-tuning with domain-specific instruction datasets. This enables geoscientists to explore a wide array of topics, generate hypotheses, and interpret complex datasets more effectively. Due to their scale and sophistication, foundation models possess the potential to accurately predict and forecast intricate scientific phenomena, such as new particle formation (NPF) and cyclone events, across varied geographical regions. By addressing both atmospheric processes and AI's evolving challenges, these models help to drive forward advancements in both fields.

\subsubsection{Generative AI Models}  

Generative AI encompasses a class of algorithms and models designed to create new, previously unseen data that closely resembles existing examples by learning the underlying patterns and structures present in the training ata~\cite{hagos2024recent}. This category includes various model types, such as generative adversarial networks (GANs), variational autoencoders (VAEs), and diffusion models~\cite{bengesi2024advancements}. These models leverage advanced techniques to mimic statistical processes, enabling the generation of synthetic data that augments datasets where labeled data are limited, thereby aiding in the training of more robust AI models \cite{hagos2024recent}.

In atmospheric science, the GenCast model developed by Google DeepMind~\cite{price2024probabilistic} operates as a conditional diffusion model, capable of producing new samples from a given data distribution. GenCast has demonstrated its ability to generate high-resolution, probabilistic weather forecasts by learning from historical climate data. By producing ensemble simulations that capture uncertainty and rare atmospheric phenomena, GenCast surpasses conventional numerical weather prediction (NWP) methods, enhancing disaster prediction and mitigation capabilities. More generally, while diffusion models have primarily focused on web content such as text, images, and video, they hold significant potential for bridging multi-modal datasets used in atmospheric sciences. 

A key advantage of generative AI is its effectiveness in addressing the challenge of limited labeled data, making it particularly adept at supporting unsupervised and semi-supervised learning. By generating synthetic data that replicates real-world conditions, these models improve performance when labeled datasets are scarce. This characteristic is especially valuable in semi-supervised learning frameworks, where small amounts of labeled data are supplemented with large volumes of unlabeled data. Notable applications of generative AI in atmospheric sciences include extreme weather prediction~\cite{racah2017} and hyperspectral image classification in remote sensing~\cite{wang2022self}. These applications highlight the transformative potential of generative models in advancing our understanding of complex atmospheric phenomena and improving predictive capabilities.

\subsubsection{Explainable AI (XAI)}  

Explainable Artificial Intelligence (XAI) encompasses a set of techniques and tools designed to make machine learning models more transparent, interpretable, and trustworthy for human users~\cite{dwivedi2023explainable}. This includes various methods, such as model-agnostic approaches like LIME (Local Interpretable Model-agnostic Explanations) and SHAP (SHapley Additive exPlanations), as well as inherently interpretable models like decision trees and linear regression. In atmospheric sciences, XAI plays a critical role in demystifying complex AI models, enabling researchers to trace model predictions back to the influencing factors and increasing trust in AI-driven insights. For instance, XAI has been employed to predict cirrus cloud formation, revealing the relationships between meteorological and aerosol conditions that affect cirrus properties~\cite{jeggle2023understanding}. This application illustrates how XAI can enhance our understanding of atmospheric processes and improve model accuracy.

XAI techniques provide insights into how models generate predictions, offering transparency for scientists and policymakers \cite{flora2024machine}. This transparency is crucial for fostering confidence in AI-driven climate models, particularly in regulatory and policy-driven environments where interpretability is essential for informed decision-making \cite{bommer2024finding}. AI becomes increasingly integrated into high-stakes atmospheric applications, such as climate modeling, weather forecasting, and disaster response, the importance of explainability and interpretability will further grow, as ensuring accountability and trust in these systems is essential.  Another important role for XAI is to help uncover relationships that could otherwise be unobservable. For example, a recent study involved the use of XAI to uncover the role of air pollutants, specifically \ch{O3} and \ch{NO2}, as important contributors to Alzheimer’s disease mortality rates~\cite{fania2024machine}. 

Interactive visualization tools powered by XAI can further enhance its benefits by enabling scientists to explore AI outputs, trace prediction pathways, and diagnose errors, thereby refining atmospheric models. By exposing the inner workings of AI systems, XAI not only enhances the credibility of forecasts but also facilitates collaborative research by allowing domain experts to question and refine model behavior.

\subsubsection{Physics-Informed Machine Learning (PIML)}  

Physics-informed machine learning (PIML) integrates domain knowledge and physical principles directly into machine learning models, enhancing their accuracy and interpretability~\cite{carleo2019machine}. This approach, also referred to as knowledge-guided machine learning (KGML), ensures that underlying physical constraints are respected during the learning process, leading to more robust and reliable predictions~\cite{karpatne2017theory}. A prominent example of PIML is the development of Physics-Informed Neural Networks (PINNs), a class of machine learning techniques designed to solve problems governed by partial differential equations~\cite{cuomo2022scientific}. PINNs have been successfully applied to emulate, downscale, and forecast weather and climate processes, demonstrating their potential to enhance both computational efficiency and predictive performance~\cite{kashinath2021physics}.  

An emerging innovation is the integration of PINN architecture with foundation models to embed physical laws into large-scale AI systems. This integration can reduce the complexity of foundation models while maintaining accuracy and generalizability. For instance, ClimaX, a foundation model developed by Microsoft for weather forecasting and climate projection, features a large and complex architecture~\cite{nguyen2023climax}. By incorporating PINNs into ClimaX, the model’s size and computational demands could be scaled down without compromising performance, offering a more efficient and interpretable approach for atmospheric and climate science. The synergy between PIML and traditional ML can help improve the performance of ML models while at the same time address the challenges posed by the complex and chaotic nature of atmospheric phenomena. By leveraging the strengths of both domains, researchers can develop more accurate and interpretable models that simultaneously reduce computational costs. Conversely, the atmospheric sciences provide a rich domain governed by clear physical laws, presenting significant opportunities to foster the future development of physics-informed machine learning models.

\subsubsection{Neurosymbolic AI for Causal Inference}  

Neurosymbolic AI, which combines neural networks with symbolic reasoning, represents a significant advancement in causal discovery and interpretation. By embedding symbolic rules into neural models, neurosymbolic AI can infer cause-and-effect relationships from observational data, thereby enhancing the reliability of AI systems used for climate modeling and environmental prediction~\cite{kukreja2024advancing}. This approach is particularly promising for detecting complex feedback loops in atmospheric systems, such as the interplay between aerosols and cloud formation.
\begin{figure*}[htbp]
    \centering
    \includegraphics[width=0.9\linewidth]{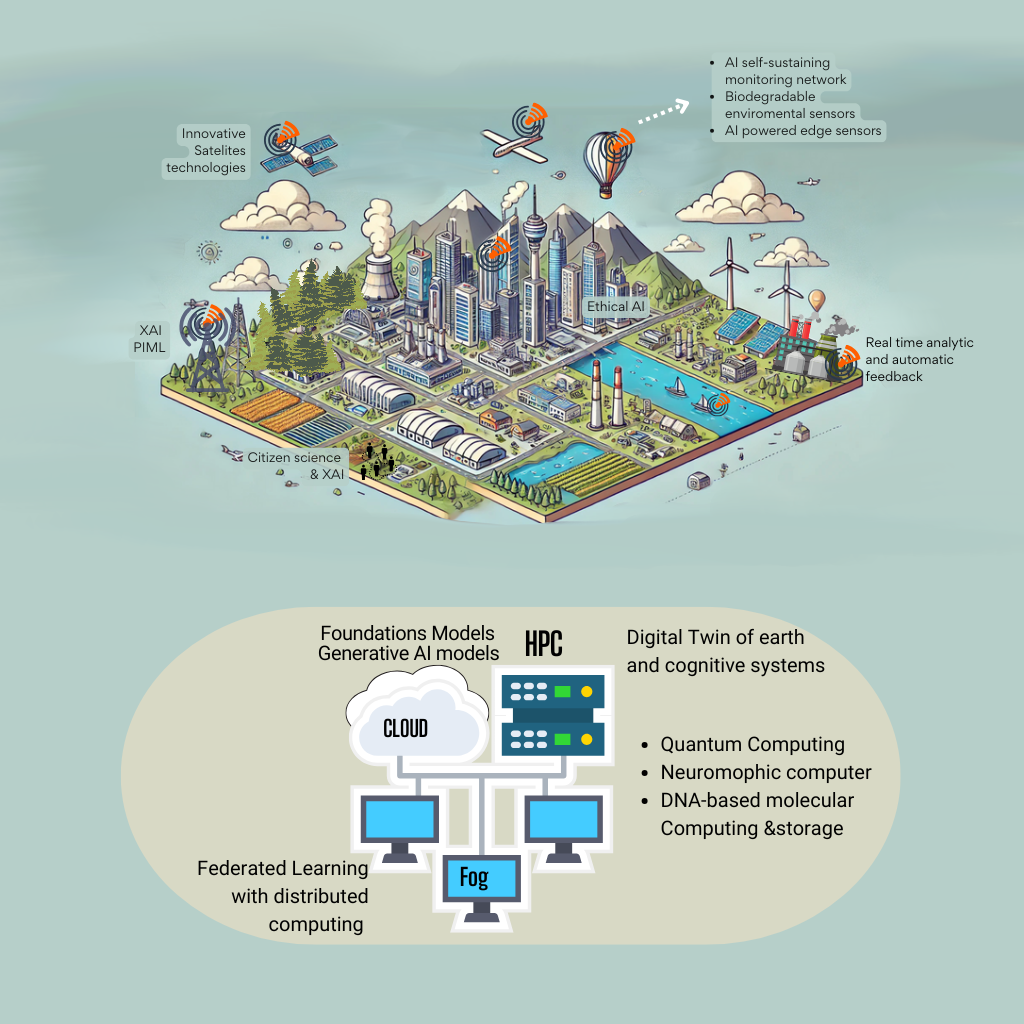}
    \caption{Emerging trends and future research directions in AI for atmospheric science and its related infrastructure}
    \label{fig:future_directions}
\end{figure*}

\subsection{From Data to Action}  

The future of atmospheric sciences hinges on ability to transform complex environmental data into actionable insights. By utilizing advanced AI-driven analytics, visualization techniques, and automated feedback systems, decision-makers can respond more effectively to environmental challenges. These innovations are crucial for identifying pressing issues and developing more effective, actionable policies that promote sustainability and resilience.

\subsubsection{Real-Time Analytics and Automated Feedback Loops}  

Real-time analytics platforms facilitate immediate processing of environmental data, and provide decision-makers with actionable insights. Adaptive systems can dynamically adjust sensor calibration processes and modify the data resolution based on current conditions. This can improve monitoring precision. AI-driven feedback loops help bridge the gap between data collection and action by autonomously controlling environmental actuators. For example, indoor air quality monitoring can adjust Heating, Ventilation, and Air Conditioning (HVAC) operations to maintain optimal conditions~\cite{pantelic2023cooking}. Similarly, flood barriers can be automatically deployed in response to real-time weather data~\cite{he2024enhancing}, while urban traffic systems can optimize traffic flows to reduce emissions~\cite{lv2023impacts}.

AI-driven analytics also enhance the management of renewable energy systems. Hyper-local weather estimation and forecasting improve the efficiency of solar and wind energy systems by optimizing energy distribution and storage~\cite{zaidan2024irmaset}. Moreover, Reinforcement Learning (RL) has demonstrated great potential in optimizing various systems by making sequential decisions in dynamic environments. For example, RL has been successfully employed to optimize energy consumption in HVAC systems, adjusting operations based on weather predictions and building occupancy patterns~\cite{zhuang2023data}. In addition, RL frameworks can dynamically allocate resources during extreme weather events, such as redirecting power to critical infrastructure during storms or floods~\cite{li2023risk}.

These applications are examples of how real-time analytics and automated feedback systems can evolve from data interpretation into proactive frameworks. By enabling timely interventions, these technologies drive improvements in sustainability and enhance environmental resilience.

\subsubsection{Digital Twin of Earth and Cognitive Systems}  

Digital twins of the Earth are sophisticated digital replicas that encompass various scales and domains of the Earth system, enabling comprehensive monitoring, forecasting, and assessment of environmental processes and the consequences of human interventions~\cite{hazeleger2024digital}. These systems empower policymakers and scientists to explore complex phenomena, such as storm evolution, and explore "what-if" scenarios for climate interventions, urban planning, and disaster response. They can facilitate data-driven decisions regarding emission control and resource management, supporting initiatives like the Earth Virtualization Engines (EVE) program, which aims to provide personalized climate knowledge to empower informed action across society~\cite{hoefler2023earth,stevens2024earth}.

Beyond global-scale assessments, digital twins can aid energy providers in designing deep mantle geothermal energy projects and assist the reinsurance industry in testing risk mitigation strategies and disaster management scenarios~\cite{bauer2021digital}. Their applications extend to agriculture, where they monitor crop health and optimize resource use. At the urban scale, digital twins can be downscaled to enhance city systems by integrating advanced visualization tools, such as augmented reality (AR) and holographic displays, optimizing air quality management, and supporting smart city initiatives~\cite{peldon2024navigating}. Despite their transformative potential, the implementation of digital twins faces challenges such as the need for high-quality data, data privacy concerns, and the complexity of accurately modeling dynamic systems. Addressing these challenges is essential for realizing the full potential of digital twins in tackling environmental and urban challenges.

\subsubsection{Citizen Science and Explainable AI for Public Engagement}  

Citizen science initiatives harness crowdsourced data from citizens, their mobile devices and diverse IoT-enabled sensor devices to augment traditional monitoring systems while fostering public awareness and participation. These initiatives enable everyday citizens to contribute to scientific research and environmental monitoring efforts. AI models, including ML algorithms, validate and process this data, ensuring reliability for decision-making. Explainable AI is also necessary to enhance the transparency regarding how models generate predictions, helping policymakers and stakeholders understand AI-driven insights and fostering trust in these systems~\cite{flora2024machine, bommer2024finding}. 
Gamified platforms are another notable method for improving public awareness. These  engage the public by simulating environmental scenarios and encouraging sustainable behavior~\cite{senka2024using}. By incorporating elements of play and competition, these platforms make participation more appealing and effective. Furthermore, citizen science can be amplified through educational workshops and community events that emphasize collective action for environmental sustainability.

\subsection{Ethical, Collaborative, and Regulatory Considerations}  

The integration of AI into atmospheric sciences presents not only technological advancements but also critical challenges in ethics, collaboration, and regulation~\cite{de2021artificial}. Addressing these aspects is essential to ensure responsible and impactful use of AI in monitoring and predictive systems.  

\subsubsection{Data Governance and Privacy}  

Atmospheric monitoring involves large-scale data collection from diverse public and private sources, raising concerns about the ethical handling of sensitive information. Ensuring data governance and privacy requires the adoption of policies that prioritize data anonymization, secure storage, and transparent usage~\cite{janssen2020data}. These measures safeguard individual and organizational privacy while maintaining data integrity~\cite{tuia2021toward}. Addressing these concerns is critical to building trust in AI-driven systems and facilitating their widespread adoption in atmospheric sciences.  

\subsubsection{Interdisciplinary Collaboration} 

The interdisciplinary nature of AI and atmospheric science integration necessitates collaboration between atmospheric scientists, AI experts, policymakers, and industry stakeholders~\cite{joshi2024artificial}. Joint efforts are essential to tackle complex environmental challenges, ensuring that AI tools are designed to meet the needs of all stakeholders. Partnerships with technology providers, regulatory agencies, and international organizations can further drive AI adoption on a global scale, creating solutions that effectively address environmental monitoring and climate change mitigation~\cite{kulkov2024artificial}.  

\subsubsection{Regulatory Frameworks and Policy Support}  

Advancements in AI require robust regulatory frameworks to ensure transparency, accountability, and ethical usage in environmental monitoring systems~\cite{diaz2023connecting}. Governments and international organizations must establish policies that support responsible AI development, enable cross-border data sharing, and ensure compliance with environmental regulations. AI technologies can also play a proactive role in policy-making by offering real-time insights and actionable recommendations, allowing policymakers to respond swiftly to environmental challenges while maintaining regulatory alignment~\cite{dwivedi2021artificial}.  

\subsection{Summary}

As our overview has consistently highlighted, the integration of advanced AI techniques into atmospheric sciences holds significant promise for enhancing our understanding of complex environmental phenomena. The proposed roadmap, illustrated in Figure~\ref{fig:future_directions}, encompasses the entire computing continuum, enhancing the analysis of atmospheric datasets and providing benefits throughout the entire data life cycle from collection to end-user applications.

In terms of deployment, the AI roadmap envisions integration across a broad computing spectrum, from low-end sensor devices to fog and edge nodes, as well as cloud and high-performance computing (HPC) platforms. Data-producing devices can leverage federated learning to ensure data privacy and sovereignty while utilizing calibration models to improve measurement quality and derive localized insights. As emerging paradigms such as DNA computing and neuromorphic computing mature, they present opportunities to further enhance computational efficiency. HPC and large-scale cloud platforms facilitate the analysis of vast datasets using foundation models and generative AI, which improve the accuracy of weather forecasts and climate models. These AI techniques excel at handling datasets comprising of multiple modalities and limited labels, common characteristics of measurements in many atmospheric domains. Looking further ahead, quantum computing has the potential to revolutionize the processing of atmospheric data, enabling the discovery of complex patterns that are currently beyond our reach. However, the successful adoption of these AI techniques hinges on building trust in AI-derived insights, necessitating the implementation of explainable AI to trace decision-making processes and the use of physical computing techniques to ensure they adhere with physical laws governing atmospheric processes.

Conversely, advancements in atmospheric sciences present unique challenges and opportunities for AI research, especially concerning the processing of multi-modal data and ensuring the robustness of AI models against variations in data characteristics. The complexity of atmospheric data sources, including satellite imagery, sensor readings, and climate models, necessitates AI systems to be capable of effectively integrating and processing heterogeneous data types. This demand pushes the boundaries of AI research, challenging current methods. As researchers work to enhance the resilience of AI models in the face of unpredictable atmospheric phenomena, they are likely to uncover new approaches that improve AI's generalizability and performance,  enriching the field of AI itself.

The implications of integrating AI into atmospheric sciences extend well beyond academic advancements. Enhanced weather forecasting and climate modeling can lead to better preparedness for extreme weather events, helping to save lives and reduce economic losses. AI-driven environmental monitoring systems can promote sustainable practices in sectors such as agriculture, energy, and urban planning, increasing societal resilience. Citizen science initiatives that utilize these technologies enhance public engagement while empowering communities to take informed actions regarding environmental issues. As we confront the complexities of climate change, the responsible integration of AI into atmospheric sciences can facilitate informed policy-making and collaborative efforts aimed at achieving environmental sustainability, helping to shape a more resilient future for both society and the planet.

\section{Conclusion} 
\label{sec:conclusion}

We examined the transformative potential AI can have for atmospheric sciences, emphasizing its potential to enhance our understanding of complex environmental systems. At the same time, we have highlighted opportunities on how atmospheric sciences can boost AI research by presenting unique challenges that push the boundaries of current methods. By leveraging measurements from diverse sources, including satellite remote sensing, ground-based measurements, and IoT devices, AI technologies are transformed our understanding of complex environmental systems and revolutionizing air quality monitoring, operational meteorology, and climate modeling. 

We identified key applications of AI across key atmospheric science domains, including air quality monitoring and modeling, in-situ atmospheric observations, operational meteorology, satellite remote sensing, Earth system modeling, and climate change. Our analysis highlighted the current state of the field, ongoing challenges, and potential solutions. While AI offers unprecedented opportunities to enhance predictive capabilities, improve climate model accuracy, and enable more efficient environmental monitoring, data scarcity, model interpretability, development of hybrid AI-physics models and other challenges must be first addressed. Looking ahead, we also presented a research roadmap that outlines key roles emerging technologies, including foundation models and quantum computing, can play for future innovations to enable the next generation of research advances in atmospheric sciences. 

To conclude, the integration of AI into atmospheric sciences not only promises to improve our understanding and management of environmental phenomena but also offers a pathway toward more sustainable interactions with our planet. By fostering collaboration between AI and atmospheric research, we can drive forward advancements that will benefit both fields and contribute to effective climate change mitigation and adaptation strategies.

\section*{acknowledgment}

This project is supported by the Research Council of Finland through the Academy Research Fellowship projects (Grant No. 355330 and 362594), an Academy Project (Grant No. 339614), and the University Profiling funding initiative InterEarth (Grant No. 353218). Additional financial support from the Research Council of Finland is gratefully acknowledged via the Atmosphere and Climate Competence Center (Grant No. 337549, 357902, and 359340), as well as through the Research Infrastructure of the Institute for Atmospheric and Earth System Research – INAR RI (Grant No. 367739). We also acknowledge support from the RI-URBANS project (Research Infrastructures Services Reinforcing Air Quality Monitoring Capacities in European Urban and Industrial Areas), funded by the European Union's Horizon 2020 research and innovation programme under the Green Deal initiative (Grant Agreement No. 101036245).

We acknowledge the use of OpenAI's ChatGPT-4 (https://chat.openai.com/) to refine the academic language and coherence of selected sections.






\bibliographystyle{IEEEtran}
\bibliography{bibtex/References}

\section*{Biographies}

\begin{IEEEbiography}[{\includegraphics[width=1in,height=1.25in,clip,keepaspectratio]{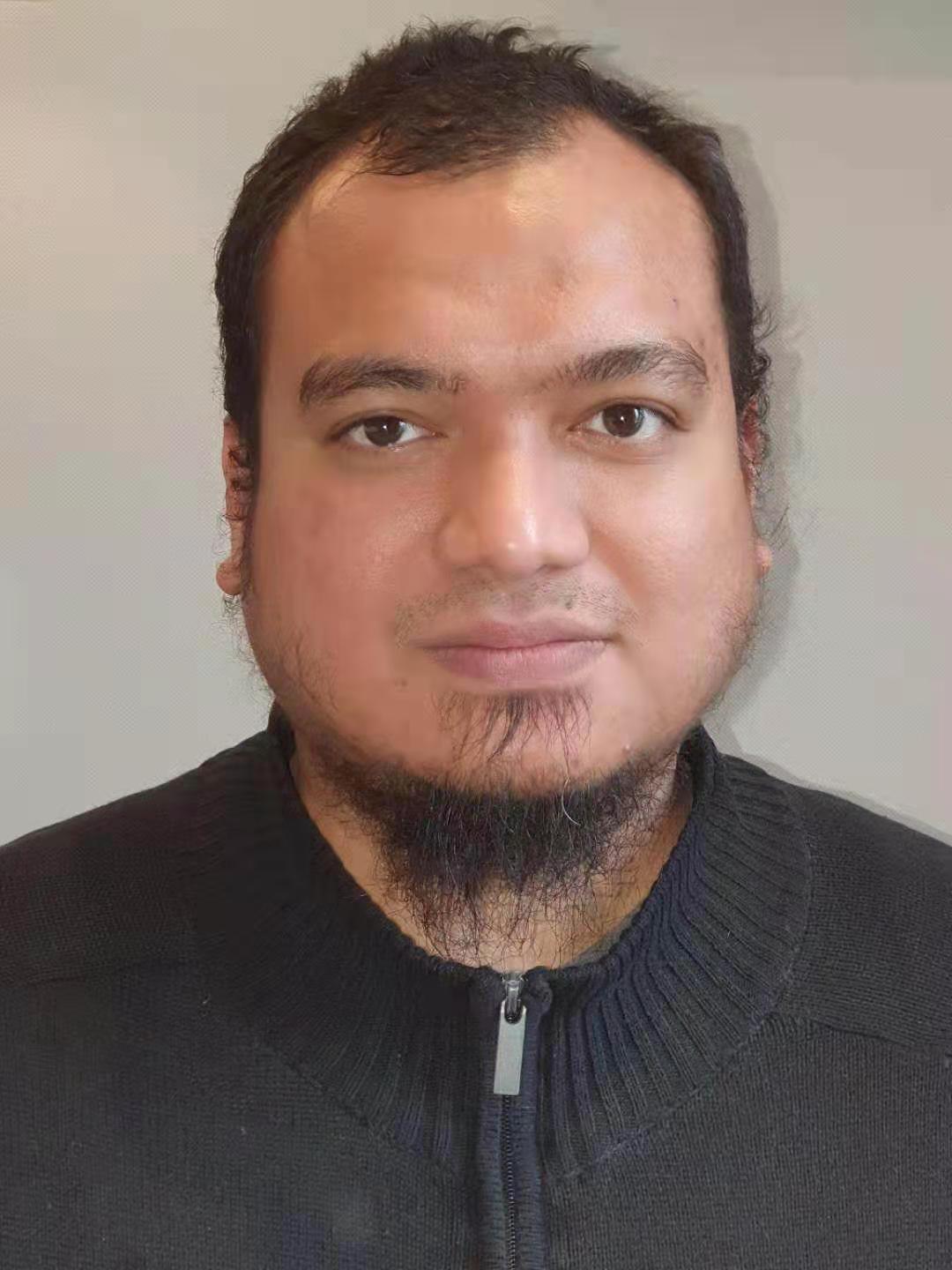}}]{Martha Arbayani Zaidan} (Senior Member, IEEE) received the Ph.D. degree in Automatic Control and Systems Engineering from the University of Sheffield. He was a Postdoctoral Research Associate at the University of Maryland, a Fellow at Aalto University, and a Research Associate Professor at Nanjing University. Currently, he is an Academy Research Fellow and Data Scientist at the Department of Computer Science and Institute for Atmospheric \& Earth System Research (INAR), University of Helsinki. His research interests include artificial intelligence and machine learning for intelligent control systems, health monitoring technologies, applied physics, and atmospheric and environmental sciences.
\end{IEEEbiography}

\begin{IEEEbiography}[{\includegraphics[width=1in,height=1.25in,clip,keepaspectratio]{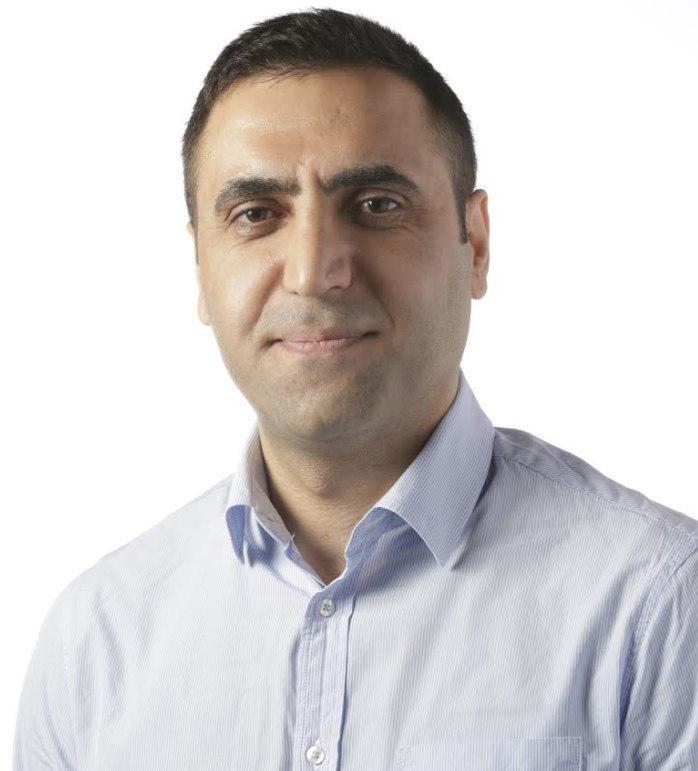}}]{Naser~Hossein~Motlagh} received the D.Sc. degree in Networking Technology from the School of Electrical Engineering, Aalto University, Finland, in 2018. He is an Academy of Finland Research Fellow  at the Department of Computer Science, University of Helsinki and at the Nokia Center for Advanced Research (NCAR). Previously, he was a Postdoctoral Fellow within the Helsinki Institute for Information Technology HIIT and Helsinki Center for Data Science (HiDATA). His research interests include the Internet of Things, wireless sensor networks, environmental sensing, smart buildings, and unmanned aerial and underwater vehicles.
\end{IEEEbiography}


\begin{IEEEbiography}[{\includegraphics[width=1in,height=1.25in,clip,keepaspectratio]{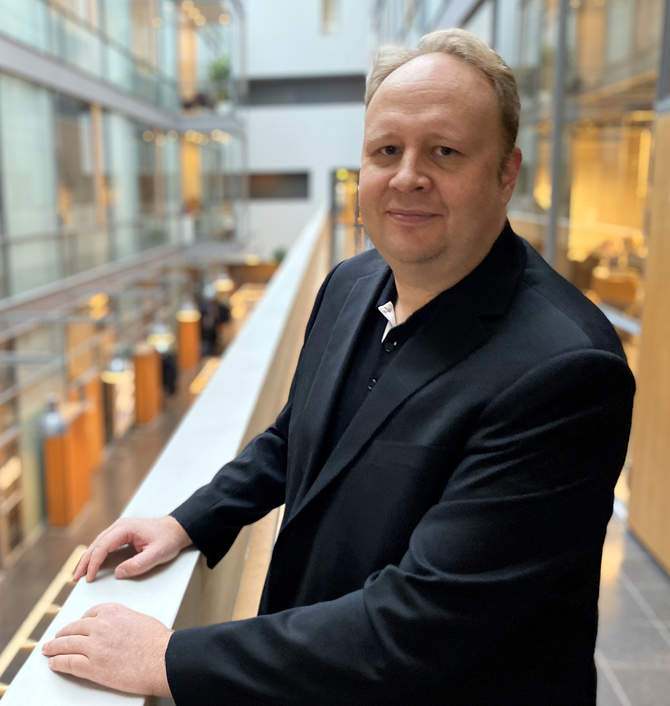}}]{Petteri~Nurmi}
is a Professor of Computer Science at the University of Helsinki, where he leads the Pervasive Data Science Group. He holds a Ph.D. (2009) and M.Sc. (2006) in Computer Science from the University of Helsinki, and was previously a Lecturer at Lancaster University.
His research focuses on pervasive computing, IoT, data science, and environmental monitoring, with a strong emphasis on intelligent mobile systems, sustainable computing, and novel sensing applications. He has published over 150 articles, including several in flagship conferences, and is highly active in international collaborations, journal publications, and conference committees.
\end{IEEEbiography}


\begin{IEEEbiography}[{\includegraphics[width=1in,height=1.25in,clip,keepaspectratio]{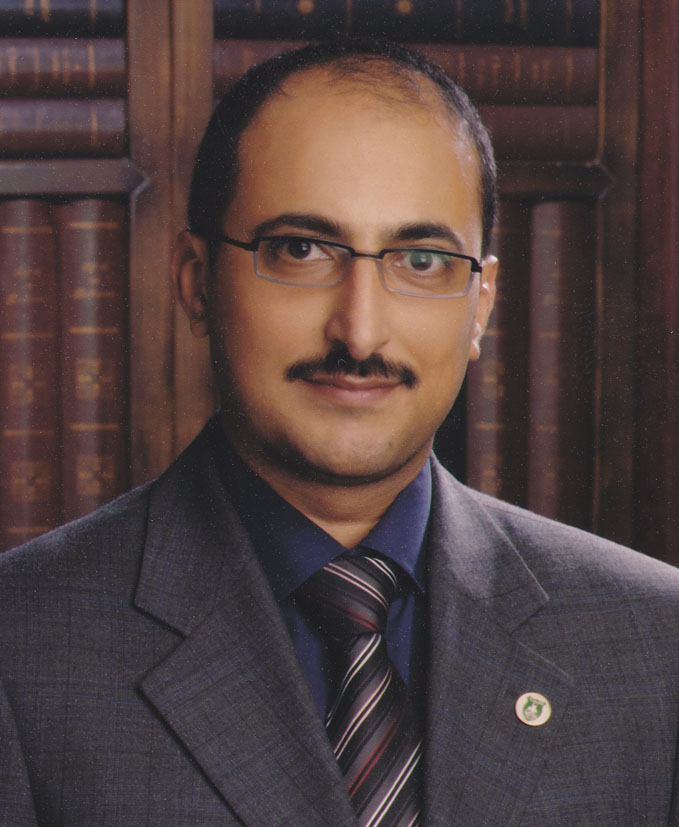}}]{Tareq~Hussein} is a Professor at the Institute for Atmospheric and Earth System Research (INAR), University of Helsinki, and a Professor at the University of Jordan. He holds a Ph.D. in Physics (2005) and is a Docent in Physics from the University of Helsinki. His research focuses on atmospheric and environmental sciences, with a strong emphasis on air pollution, urban and indoor air quality. This includes the dynamics and physical characterization of aerosol particles, emissions, dry deposition, and health effects.
\end{IEEEbiography}


\begin{IEEEbiography}[{\includegraphics[width=1in,height=1.25in,clip,keepaspectratio]{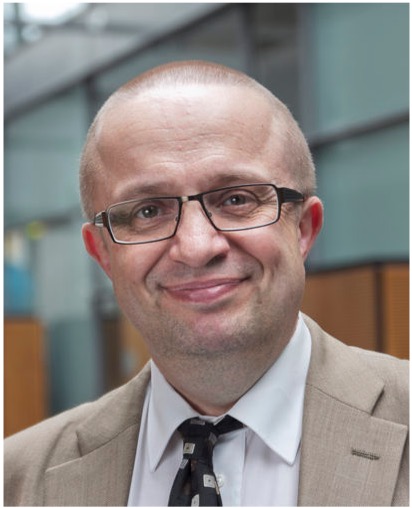}}]{Markku~Kulmala}
who leads INAR, is a highly influential researcher in atmospheric science. He's known for his work on global measurement networks and air quality-climate change interactions, particularly through SMEAR and GlobalSMEAR. With over 1100 original research papers, including 18 in \textit{Nature} and 17 in \textit{Science}, he has over 60,000 citations and an H-factor of 120. Formerly ranked first in Geosciences citations, he is a foreign member of the Chinese and Russian Academies of Sciences, a member of five other academies, and president of the European Center of International Eurasian Academy of Sciences. He has received over 10 international awards and holds nine honorary doctorates/professorships.
\end{IEEEbiography}

\begin{IEEEbiography}[{\includegraphics[width=1in,height=1.25in,clip,keepaspectratio]{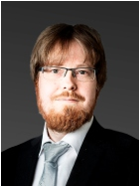}}]{Tuukka~Petäjä} is a Professor of Experimental Atmospheric Sciences and Vice-Director of INAR at the University of Helsinki. He earned his Ph.D. (2006) and Docent (2011) in Physics from the University of Helsinki and completed postdoctoral research at the U.S. National Center for Atmospheric Research (NCAR). He oversees INAR's aerosol domain and research infrastructures. He has over 430 peer-reviewed articles, including 10 in Science and 9 in Nature, with over 22,000 citations and an H-factor of 69. Recognized as a highly cited scientist since 2014, his awards include the FAAR Award, Vaisala Award (2013), and Science and Technology in Society Future Leader Award. He also leads the BAECC campaign, directs PEEX, serves on the PACES initiative board, and is an academician in the International Eurasian Academy of Sciences.

\end{IEEEbiography}

\vspace{-0.4cm}

\begin{IEEEbiography}[{\includegraphics[width=1in,height=1.25in,clip,keepaspectratio]{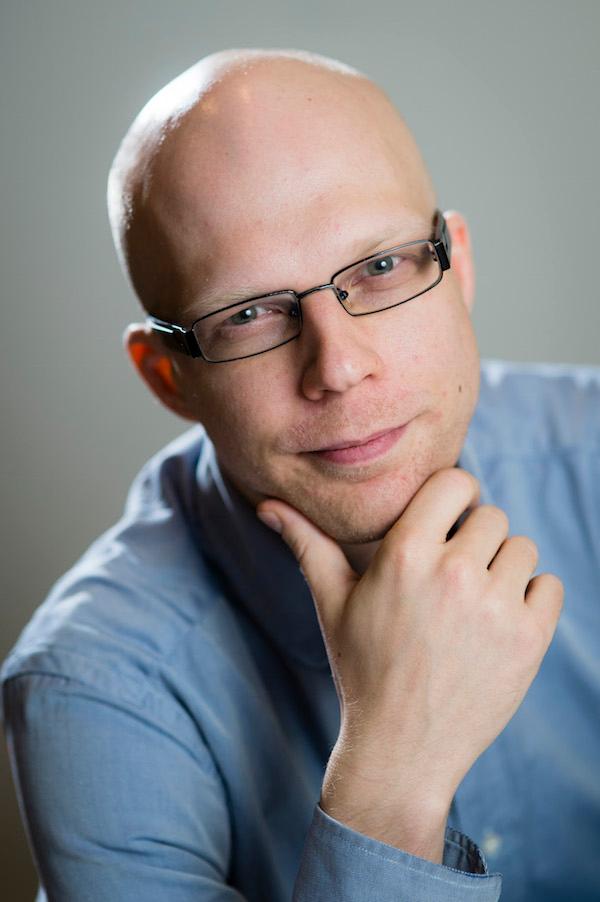}}]{Sasu~Tarkoma} (Senior Member, IEEE) is  a Professor of Computer Science and Dean of the Faculty of Science at the University of Helsinki. His research focuses on edge intelligence, IoT, and distributed systems. He has 472 publications, 14,463 citations, and an H-index of 57 (AD Scientific Index 2025). Author of 4 textbooks, he holds 11 US patents and over 20 international patent applications. Professor Tarkoma has received numerous Best Paper awards (IEEE PerCom, ACM CCR, ACM OSR) and chairs the Finnish Scientific Advisory Board for Defence (MATINE). He is a Fellow of IET and EAI.
\end{IEEEbiography}




\end{document}